\documentclass[preprint,12pt,numbers,sort&compress]{elsarticle}
\usepackage{lineno}
\usepackage{amssymb}
\usepackage{footmisc}
\usepackage{array}
\usepackage{graphicx}
\usepackage{amsmath}
\usepackage{hyperref}
\usepackage[T1]{fontenc}
\hypersetup{
  colorlinks=true,
  linkcolor=blue,
  urlcolor=blue,
  citecolor=blue
}
\usepackage{booktabs}
\usepackage{tabularx}
\usepackage{caption}
\usepackage{url}
\usepackage{longtable}
\usepackage{multirow} 
\usepackage[ruled,vlined]{algorithm2e}
\usepackage[a4paper, left=1in, right=1in, top=1in, bottom=1in]{geometry}
\usepackage{algorithmic}
\usepackage{setspace} 
\usepackage{float} %
\usepackage{pdflscape}
\usepackage{pifont}

\journal{Physics Report}
\makeatletter

\def\ps@pprintTitle{
  \let\@oddhead\@empty
  \let\@evenhead\@empty
  \let\@oddfoot\@empty
  \let\@evenfoot\@empty
}

\patchcmd{\pprintMaketitle}{\vskip36pt}{\vskip15pt}{}{}

\makeatother

\begin{document}
\onehalfspacing
\begin{frontmatter}

\title{{\LARGE\textbf {Human Mobility in Epidemic Modeling}}}

\author[inst1]{\raggedright Xin Lu\corref{cor1}\fnref{equal}}
\author[inst1,inst2,inst3,inst4]{Jiawei Feng\corref{cor1}\fnref{equal}}
\author[inst5]{Shengjie Lai}
\author[inst6,inst7]{Petter Holme}
\author[inst1]{Shuo Liu}
\author[inst8,inst9]{Zhanwei Du}
\author[inst1]{Xiaoqian Yuan}
\author[inst2]{Siqing Wang}
\author[inst1]{Yunxuan Li}
\author[inst1]{Xiaoyu Zhang}
\author[inst8,inst9]{Yuan Bai}
\author[inst10]{Xiaojun Duan}
\author[inst2]{Wenjun Mei}
\author[inst11]{Hongjie Yu}
\author[inst1]{Suoyi Tan\corref{cor1}}
\author[inst12]{Fredrik Liljeros}

\address[inst1]{\raggedright\textit{College of Systems Engineering, National University of Defense Technology, Changsha, China}}
\address[inst2]{\raggedright\textit{College of Engineering, Peking University, Beijing, China}}
\address[inst3]{\raggedright\textit{Department of Real Estate and Construction, The University of Hong Kong, Hong Kong SAR, China}}
\address[inst4]{\raggedright\textit{Sustainability X-Lab, The University of Hong Kong, Hong Kong SAR, China}}
\address[inst5]{\raggedright\textit{WorldPop, School of Geography and Environmental Science, University of Southampton, Southampton, UK}}
\address[inst6]{\raggedright\textit{Department of Computer Science, Aalto University, Espoo, Finland}}
\address[inst7]{\raggedright\textit{Center for Computational Social Science, Kobe University, Kobe, Japan}}
\address[inst8]{\raggedright\textit{World Health Organization Collaborating Center for Infectious Disease Epidemiology and Control, School of Public Health, Li Ka Shing Faculty of Medicine, The University of Hong Kong, Hong Kong SAR, China}}
\address[inst9]{\raggedright\textit{Laboratory of Data Discovery for Health Limited, Hong Kong Science and Technology Park, Hong Kong SAR, China}}
\address[inst10]{\raggedright\textit{College of Science, National University of Defense Technology, Changsha, China}}
\address[inst11]{\raggedright\textit{Department of Epidemiology, School of Public Health, Key Laboratory of Public Health Safety, Ministry of Education, Fudan University, Shanghai, China}}
\address[inst12]{\raggedright\textit{Department of Sociology, Stockholm University, Stockholm, Sweden}}

\fntext[equal]{These authors contributed equally to this work.}
\cortext[cor1]{Corresponding authors. Emails: xin.lu.lab@outlook.com (XL), fengjiawei126@gmail.com (JF), tansuoyi\_cn@outlook.com (ST).}

\begin{abstract}
Human mobility forms the backbone of contact patterns through which infectious diseases propagate, fundamentally shaping the spatio-temporal dynamics of epidemics and pandemics. While traditional models are often based on the assumption that all individuals have the same probability of infecting every other individual in the population, a so-called random homogeneous mixing, they struggle to catch the complex and heterogeneous nature of real-world human interactions. Recent advancements in data-driven methodologies and computational capabilities have unlocked the potential of integrating high-resolution human mobility data into epidemic modeling, significantly improving the accuracy, timeliness, and applicability of epidemic risk assessment, contact tracing, and intervention strategies. This review provides a comprehensive synthesis of the current landscape in human mobility-informed epidemic modeling. We explore several data sources and representations of human mobility, and examine the behavioral and structural roles of mobility and contact in shaping disease transmission dynamics. Furthermore, the review spans a wide range of epidemic modeling approaches, ranging from classical compartmental models to network-based, agent-based, and machine learning models. It also discusses how mobility integration enhances risk management and response strategies design during epidemics. By synthesizing these insights, the review can serve as a foundational resource for researchers and practitioners, bridging the gap between epidemiological theory and the dynamic complexities of human interaction while charting clear directions for future research.
\end{abstract}

\end{frontmatter}

\tableofcontents
\listoffigures
\listoftables

\section{Introduction}\label{sec:1}

A disease outbreak or epidemic, as defined by the World Health Organization (WHO) \citep{2025v}, is a sudden increase in incidence beyond what is usually anticipated within a specific population, community, geography, or season. These outbreaks are often driven by the spread of infectious agents through various pathways, including direct person-to-person contact, animal reservoirs, environmental exposure, or vector-borne mechanisms involving insects and animals \citep{Cohen2000,Heesterbeek2015,Hossain2022,Kraemer2021,Lloyd-Smith2009}. At the core of these transmission routes lies human behavior, particularly the movement and interactions of individuals, which fundamentally shapes the dynamic contact structures that significantly influences where, when, and how pathogens spread \citep{Kostandova2024a,Lloyd-Smith2009,Lu2012,Mistry2021,Perofsky2024,Sun2025c,Viboud2006,Zhang2023e,Zhong2025}. Among behavioral drivers, human mobility plays a particularly critical role in the disease transmission dynamics. The sudden and large-scale movement of people, whether due to travel, migration, or social congregation, can amplify localized outbreaks into regional epidemics or even global pandemics \citep{Jia2020}. When infected individuals move from areas of high to low disease prevalence, they probably introduce novel risk into previously unaffected populations or disease-eliminated regions. These dynamics underscore the importance of understanding and modeling mobility accurately in epidemiological frameworks for infectious diseases, to support timely epidemic intervention and mitigation.

Epidemic modeling has undergone significant advancements since its inception. The foundations of quantitative epidemiology were established through the statistical analysis of mortality data and further advanced by applying mathematical methods to evaluate the impact of smallpox vaccination \citep{Dietz2002}. Thereafter, a significant milestone was the development of the SIR (Susceptible-Infected-Recovered) model, which simplified the complex dynamics of diseases spread by categorizing populations into specific compartments \citep{Kermack1927}. These models, including SIR and its variations like SIS (Susceptible-Infected-Susceptible) and SEIR (Susceptible-Exposed-Infected-Recovered), not only simulate disease progression but also facilitate the derivation of key analytical indicators that characterize epidemic potential \citep{Chinazzi2020,Colizza2007,Peak2018,Poletto2014,Wesolowski2014,Yang2009}. Foremost among these is the basic reproduction number \citep{Liu2018}, $R_0$, which quantifies the excepted number of secondary infections caused by a single infectious individual in a fully susceptible population.

However, traditional compartmental models rest on the assumption of homogeneous mixing, where each individual is equally likely to contact any other. Although some are analytically tractable (e.g., SI), they remain a significant simplification of real-world contact structures \citep{Cozzo2013,DAngelo2023,Hossain2022,Lu2025b,Murayama2024}. Real populations are spatially distributed, socially stratified, and temporally heterogeneous \citep{Colizza2007a}. As such, contact rates vary over time and space, modulated by factors such as daily commuting, international travel, social distancing policies, and cultural practices \citep{Broekaert2024}. To account for these complexities, epidemic models have been extended into metapopulation structures \citep{Colizza2007a}, network-based models \citep{Selinger2021}, and agent-based simulations \citep{Kerr2021}, where heterogeneous contact patterns are explicitly encoded, and mobility between subpopulations can be modeled stochastically. These methods also help estimate the effective reproduction number \citep{Trevisin2023}, $R_e(t)$, which reflects the average number of new infections per case at time $t$, under prevailing immunity levels, health behavior, and interventio  n measures. When ${R_e} < 1$, the epidemic is expected to wane; otherwise, it stands (${R_e} = 1$) or grows (${R_e} > 1$).

Beyond average-case dynamics, individual-level heterogeneity in transmission may play a pivotal role in amplifying local outbreaks.  Certain individuals due to higher connectivity, elevated viral load, or extensive mobility infect many others disproportionately \citep{Holme2019,Song2010}. This overdispersion in secondary infections violates the assumptions of deterministic models and necessitates stochastic modeling approaches, particularly in low-prevalence settings where chance events dominate \citep{Chen2024a}. The degree of overdispersion is often characterized using negative binomial distributions and has critical implications for outbreak predictability and control strategies \citep{Oztig2020}.

\begin{figure*}[htbp]
\centering
\includegraphics[width=\textwidth]{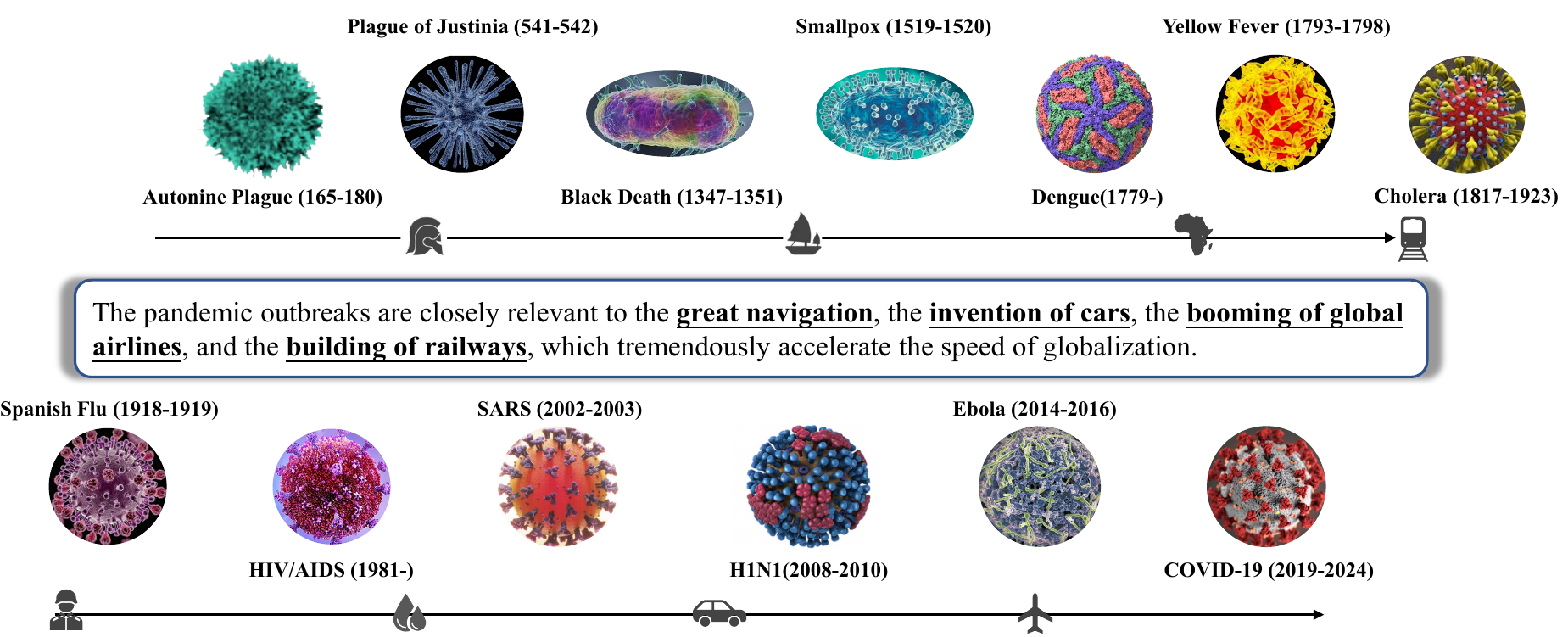}
\caption{Chronological outline of major pandemics and epidemic events in human history.}

{\small\noindent
    \parbox{\textwidth}{
      The timeline demonstrates when representative infectious diseases emerged or peaked (from the Antonine Plague in approx.\ 165--180 CE to the COVID-19 pandemic in approx.\ 2020--2023), together with their causative pathogens. The icons at the bottom (people, ship, railway, automobile, airplane, etc.) represent successive waves of technological innovations in transportation and mobility that reshaped the speed and scope of pathogen transmission across regions. This Figure also highlights that these major historical outbreaks were closely associated with profound changes in human mobility.
    }
  }

\label{Fig.1}
\end{figure*}

Historically, major widespread epidemics have often coincided with advances in transportation and increase in global connectivity such as maritime expansion, industrialization, modern air travel, world wars and mass migrations, which have facilitated the movement of people and, by extension, pathogens \citep{Brockmann2013,Lecocq2020}. \autoref{Fig.1} demonstrates a timeline of major pandemics and infectious disease outbreaks throughout history, showing the types of pathogens responsible involved and the temporal context. Although classic epidemic models have provided valuable insights into disease dynamics, there is growing emphasis on explicitly incorporating mobility and social contact patterns, to more accurately represent complex, large-scale, and heterogeneous transmission dynamics across space and time \citep{Barbosa2018,Liu2023d}. For example, estimating time-dependent transmission metrics, such as the instantaneous reproduction number at time $t$, $R_t$, has become essential for assessing near real-time epidemic risk and intervention effectiveness. These estimates are typically inferred through Bayesian filtering or likelihood-based methods, incorporating diverse observational data streams including case counts, hospitalization records, mortality data, and sometimes mobility surveys or wastewater surveillance \citep{Hinch2021,Kendall2024,Kerr2021,Overton2022}. 

The addition of mobility data, especially from anonymized mobile phone signals, transportation usage logs, and geo-referenced digital footprints, further enables the ability of models to dynamically simulate contact behaviors, pathogen dissemination patterns, spatiotemporal transmission dynamics, and the effects of targeted interventions \citep{Hu2021b,Pepe2020,Yabe2022,Zhang2022}. Advances in modern transportation and technologies have markedly enhanced the precise and real-time tracking of human movement, leading to substantial improvements in data collection and application \citep{Cauchemez2008}. Previous studies have shown that mobility data is a more effective predictor for epidemic transmission than other indicators such as web search trends, population size, or city GDP \citep{Jia2020}. Moreover, studies demonstrate that spatial proximity measures, such as distance from the outbreak source, offer limited predictive value compared with mobility-derived indicators \citep{Chen2024d,Zhang2024e}. These findings clearly demonstrate that the transmission of infectious agents is primarily driven by human movement patterns, thereby diminishing the explanatory power of static or aggregate metrics.

During the early stages of the COVID-19 pandemic \citep{Du2020a}, researchers identified a strong correlation between population movement and outbreak trajectories across communities, cities, countries and continents \citep{Morens2020}. Similar insights have emerged from earlier epidemics or pandemics, like H1N1 influenza \citep{Yang2009}, cholera \citep{Bengtsson2015}, dengue \citep{Bomfim2020}, Mpox \citep{Murayama2024}, and Ebola \citep{Malvy2019}, each of which have demonstrated how mobility patterns uniquely shape transmission dynamics based on diverse mobility modes, from daily commuting and seasonal migration to international travel. Case studies have demonstrated that data on individual travel paths and contact patterns is critical for accurate epidemic modeling \citep{Bomfim2020,Chinazzi2020,Jia2020}. For instance, during the H1N1 outbreak, global air travel patterns were critical in predicting large-scale disease spread within a short period, while for cholera and dengue, localized community movement and interactions played a larger role \citep{Stein2009}. This fusion of physics-informed network modeling, high-resolution spatiotemporal mobility patterns, and data-driven statistical inference marks a shift toward quantitative, adaptive epidemiology, where control policies are informed not just by pathogen biology but by the evolving structure of human interactions \citep{Laxminarayan2020,Quach2025}. The spread of diseases such as COVID-19 and influenza has been shown to correlate strongly with international and domestic mobility patterns, as these movements help introduce infections to new, previously unaffected regions \citep{Poletto2013}. By tracking human movement patterns, public health officials can predict and manage the spatial spread of an epidemic, especially when mobility data is combined with real-time health surveillance \citep{Petersen2020}. This combination of mobility data and health metrics allows authorities to forecast where future outbreaks are most likely to occur and direct resources to these high-risk areas \citep{Petersen2020,Poletto2013,Germann2006,Kraemer2017}.

This review aims to summarize the data sources and representations of human mobility and discuss how to incorporate them into epidemic modeling, risk management, and response strategies design. In \autoref{sec:2}, we provide an overview of mainstream data sources of human mobility relevant to past infectious disease outbreaks and future pandemic control. In \autoref{sec:3}, we introduce four representations of human mobility. They provide convenient ways to describe human mobility patterns, enabling researchers to select the most suitable format for constructing epidemic models. In \autoref{sec:4}, we explore the relationship between human mobility and epidemiology, focusing on determining whether and to what extent population flows, and contact behaviors contribute to the spread of infectious diseases. In \autoref{sec:5}, we discuss widely adopted epidemic modeling approaches integrated with human mobility, including compartmental models, network-based models, agent-based models, and machine learning models. In \autoref{sec:6}, we examine the role of human mobility in epidemic risk management, exhibiting the potential of human mobility to identify high-risk areas, track infection pressure, and enhance intervention effectiveness, thereby enabling timely risk management. In \autoref{sec:7}, we provide an overview of relevant public health policy implemented during pandemics, covering the disruption of epidemic transmission and mobility-oriented non-pharmaceutical interventions (including social distance control and lockdowns). In \autoref{sec:8}, we conclude with an outlook on the role of human mobility in epidemic modeling, highlighting methodological limitations, unresolved challenges, research gaps, and directions for future research.

\section{Human Mobility Data Sources}\label{sec:2}

\subsection{Social Surveys}\label{sec:2.1}
Social surveys have long been a foundational method of data collection in epidemiology, providing insights into public behaviors, attitudes, and demographic characteristics \citep{Chen2024d,Prem2017}, with ongoing advancements in data collection methods continually enhancing their value and applicability \citep{Pascart2024}. Through structured questionnaires or interviews, social surveys enable the collection of both qualitative and quantitative data, allowing researchers to obtain detailed, individual-level demographic and socioeconomic information from a wide range of populations \citep{Newson2024,Rao2020}. In the context of public health and epidemic modeling, social surveys through questionnaires and follow-up studies are essential for capturing human behavior, mobility patterns, and social interactions that are critical to understand disease transmission dynamics \citep{Camitz2006,Mossong2008}. For instance, survey-based estimates of contact frequency and mixing patterns have been instrumental in refining key epidemiological parameters such as the effective reproduction number \citep{Trevisin2023}. Within compartmental models (e.g., SIR frameworks, see \autoref{sec:5.1}), survey data inform transmission and contact rate, while in agent-based models, they supply granular details on individual attributes to make simulations more realistic.

Travel surveys, for instance, can reveal key details such as individuals' mobility trajectory, trip purposes, and the environmental contexts of visited locations (e.g., indoor or outdoor settings), while also including groups often underrepresented in digital datasets, such as young children without mobile phones or residents in low-connectivity, impoverished regions \citep{Eubank2004}. Meanwhile, the emergence of digital survey tools has expanded both the scope and accessibility of traditional survey data collection, motivating real-time responses and improving coverage of previously hard-to-reach groups (e.g., individuals living with AIDS) \citep{Newson2024,Quer2021}.

Despite their strengths, social surveys face several challenges. Issues such as sampling bias \citep{Haddad2022}, high non-response rates \citep{Molenberg2021}, and limitations in questionnaire design \citep{Yang2021} can significantly affect the accuracy and representativeness of the collected data. Moreover, ensuring the confidentiality and ethical handling of sensitive personal information remains a critical concern in the survey process.

\subsection{Public Transportation Records}\label{sec:2.2}
As a traditional source of human mobility data, public transportation records have long been used to acquire human movement patterns \citep{Wu2020c,Balcan2009}, including: (1) passenger transit card swipe records; (2) data autonomously collected at toll stations and gates; and (3) ticket sales and passenger transport records from vehicles, ships, airplanes, and other modes of transportation. Passenger transit card swipe records provide detailed insights into individual travel patterns within public transit systems \citep{Musa2023}, while data collected at toll stations and gates offer information on vehicular movement and traffic flows. Moreover, ticket sales and passenger transport records from various transportation modes provide aggregated data on traveler numbers and routes taken \citep{Watts2020,Zhang2014}.

Public transportation data offers valuable multi-scale insights into human mobility patterns, ranging from local commutes to international travel. These data are essential for applications in urban planning, regional development, and public health, especially in tracking and managing the spread of infectious diseases. One widely adopted pre-processed representation of public transportation records is the mobility network (see \autoref{sec:3.2}), which aggregates flows into graph-based structures for efficient analysis and epidemic modeling in the context of pandemics.

At the intra-city level, public transport systems such as buses and metros support daily commutes and short-distance travel within urban areas. These systems are heavily utilized, with daily passenger volumes reaching millions in large cities, service frequencies ranging from a few minutes during peak hours to 30 minutes during off-peak times, and geographical coverage spanning 50 to 1,000 square kilometers. Inter-city transportation \citep{Tan2021a}, primarily served by trains and long-distance buses, often connects urban centers and facilitates medium-range travel across administrative regional boundaries. This mobility data is crucial for deconstructing regional connectivity and population flow between cities, with passenger volumes varying from thousands to hundreds of thousands per day. Service frequencies vary from multiple departures per hour on popular routes to just a few per day on less-traveled ones, with geographical coverage spanning several hundred to over a thousand kilometers. International traffic \citep{Balcan2010}, primarily facilitated by air travel and ferries, plays a crucial role in understanding global connectivity and the spread of epidemics. This sector spans international routes that link countries across thousands of kilometers. Major transportation hubs handle daily passenger volumes ranging from thousands to over a million, with service frequencies varying from multiple departures per day at busy airports to weekly services on less-traveled routes.

\subsection{Cellular Signaling}\label{sec:2.3}
Cellular Signaling Data (CSD), including both active and passive signaling data, is generated when a mobile phone connects to a Base Transceiver Station (BTS) during communication activities, such as powering on, making calls, and accessing the Internet. It includes a variety of information, such as call records, text messages, internet usage, and, most importantly for epidemic modeling, location data derived from the interaction between mobile devices and BTS \citep{Csaji2013,Ye2020}. Large-scale mobile phone data, at the national level, provides a finer-grained and high-quality characterization of human activities at unprecedented resolution and scale. As such, it is an increasingly valuable data source for enhancing epidemic preparedness and response efforts. With advancements in telecommunication technology and the expanding coverage of BTS, the reach and quality of cellular networks have increased significantly. Call Detail Record (CDR) is one type of CSD, which contains information about the time of a call and the cell tower to which the mobile phone was connected when the call occurred \citep{DeMonasterio2016,Eagle2009,Oliver2020}. In detail, CDR includes the precise time and date of each transaction, an anonymized yet distinctive identifier for both the calling and receiving parties, the call duration, and details of the cellular towers involved in the call (see \autoref{tab.1}).

\begin{table}[htbp]
  \caption{An example of mobile phone signaling data fields.}
{\small\noindent
    \parbox{\textwidth}{
  The table shows the anonymous user ID, interaction time, location, and base station code. According to the information, researchers can easily trace the users’ spatiotemporal trajectories.
    }
  }
  \label{tab.1}
  \begin{center}
  \begin{tabular}{ll}
  \toprule
  \textbf{Fields} & \textbf{Examples} \\
  \midrule
  Time & 20240210000040 \newline (02/10/2024 0:0:40 PM) \\
  User ID & 99168999959434191 \\
  Phone number & 1506179 (Top seven numbers) \\
  User behaviour & RAU-NORMAL \\
  Location code & 20512 \\
  Base station code & 12321 \\
  \bottomrule
  \end{tabular}
  \end{center}
  \end{table}

In the context of pandemics, billions of mobile users generate a vast amount of CSD that can be used to track human movements. These CSD are typically aggregated (e.g., counties and states) for public release or further processed into index-based data to represent human mobility at the population level \citep{Hu2021b}. Population mobility data from BTS is often incomplete due to multiple operators within a country. For instance, China has China Mobile, China Unicom, and China Telecom, while the U.S. has AT\&T, Verizon, and T-Mobile. Therefore, these data often need to be pre-processed using machine learning algorithms to infer and model population mobility trends. 

During the COVID-19 pandemic, CSD demonstrated significant potential in enhancing outbreak response strategies \citep{Qi2023,Shin2016}, while several limitations also became apparent. Notably, datasets from major urban centers tend to be more reliable than those from rural areas, leading to an urban bias that limits representativeness. As a result, mobility patterns in less populated regions—often where healthcare resources are scarcest—may be overlooked \citep{Haddad2022}. Furthermore, concerns about privacy and restrictions on data accessibility hinder the widespread sharing and integration of CSD \citep{Wieringa2021}. Addressing these challenges is critical to improving the accuracy and equity of epidemic modeling and ensuring the effectiveness of interventions across diverse populations.

\subsection{Satellite Positioning}\label{sec:2.4}
Satellite positioning refers to the use of satellite systems to determine precise geographic locations on the Earth's surface. The most widely recognized satellite positioning system is the Global Positioning System (GPS), developed by the United States. Besides, several other systems exist, and each serves similar functions: BeiDou (BDS) from China, Galileo from the European Union, and GLONASS from Russia. These systems collectively provide comprehensive global coverage, enabling fine-grained location tracking. Satellite positioning technologies have emerged as essential tools for tracking human mobility. Numerous companies collaborate with mobile app operators to gather user satellite positioning data through app agreements, with user consent. The availability of such datasets significantly aided policy-making during the early stages of pandemics \citep{Zhang2023e}.

\begin{figure*}[htbp]
\centering
\includegraphics[width=\textwidth]{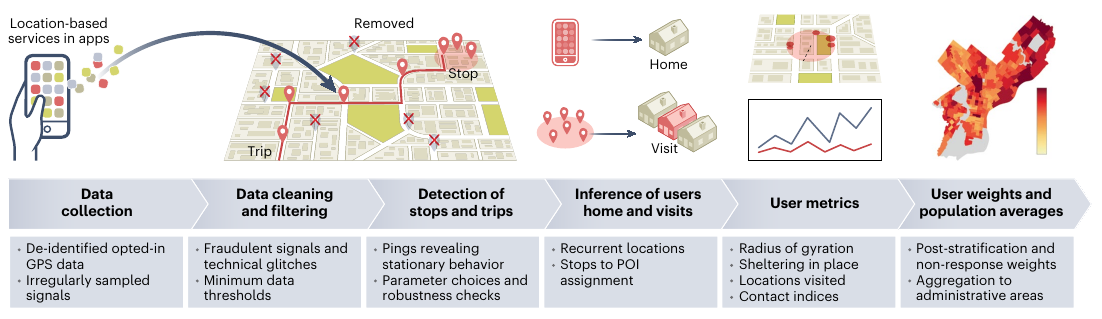}
\caption{Broad overview of anonymized GPS data applications.}

{\small\noindent
    \parbox{\textwidth}{
 The diagram summarizes the pipeline from opt-in, de-identified smartphone GPS traces to population-level mobility metrics used in epidemic analysis. Raw, irregularly sampled signals are first collected from location-based services and then cleaned and filtered to remove noise, fraudulent pings, and technical artifacts. Processed trajectories are used to detect stops and trips, allowing the identification of home locations and visits to points of interest (POIs) from recurrent spatial patterns. On this basis, user-level indicators are derived. Finally, post-stratification and weighting procedures correct sampling bias and enable aggregation to administrative regions, producing statistically representative mobility measures for epidemiological modeling. This figure is a reproduction of Fig. 1 in Ref. \citep{Barreras2024}.}
}
\label{Fig.2}
\end{figure*}

The prevalence of telecommunication devices, particularly smartphones, enables the reveal of an individual’s daily mobility behaviors through GPS traces, such as where and how long they remain in a particular location \citep{Barreras2024,Zhao2016}. In the context of epidemics, GPS data can support the identification of contact events, assess mobility reduction during interventions, and help construct contact networks to simulate spatial spread of infections. However, its high resolution also raises privacy concerns, requiring strict anonymization and user consent measures \citep{Hu2021b}. For instance, some applications request access to the smartphone’s GPS function in order to upload location data to cloud services and provide relevant functionalities. These data will be stored and analyzed in an anonymized form. \autoref{Fig.2} demonstrates the broad overview of anonymized GPS data applications, which includes data collection, cleaning and filtering, detection, inference, metrics construction, and visualization \citep{Barreras2024}.

\subsection{IP and Wi-Fi Location Tracking}\label{sec:2.5}
Internet Protocol (IP) addresses can reveal the approximate geographic location of devices based on their network connections, aiding in the tracking of movements across different regions \citep{Yook2002}. Wi-Fi data, combined from multiple access points, can provide even more precise information about an individual's location according to signal strengths, especially within buildings or urban areas. When devices connect to public networks, they interact with access points that log the device’s presence, allowing inferences about the location and movement of users within the range of these access points. Wi-Fi data is instrumental in indoor environments where satellite positioning signals may be weak or unavailable. Each device connected to the internet will be assigned an IP address that can be used to approximate its geographic location. Although IP addresses are less precise than GPS or Wi-Fi data, they offer accessible information on mobility at the regional and city levels. This data is commonly leveraged by internet service providers to provide roughly estimated human mobility in pandemics. 

During the COVID-19 pandemic, Wi-Fi and IP address data proved useful for monitoring crowd density and mobility patterns, as well as assessing the effectiveness of lockdown measures by tracking reductions in foot traffic within public spaces (see \autoref{sec:7.2}) \citep{Reymondin2024}. These data sources play a significant role in the broader human mobility data landscape, complementing other sources by providing additional detail and helping to fill gaps where GPS or other tracking methods may be less effective. Their integration supports a range of applications, including behavioral analysis and epidemic modeling.

\subsection{IoT Location Tracking}\label{sec:2.6}

The Internet of Things (IoT), a vast network of interconnected devices exchanging data \citep{Hernandez-Orallo2020,Xu2022}, offers a rich source about human mobility. These mobility patterns can be inferred not just from common wearables like smartwatches, but from a diverse ecosystem that includes fitness trackers providing granular GPS traces, smartphones logging Wi-Fi and Bluetooth beacon interactions, and connected vehicles reporting their movement \citep{Mirjalali2022}. The core technical opportunity lies in the fusion of this heterogeneous data, which, when aggregated, can create a high-fidelity view of individual and collective movement over time \citep{Rosa2021}.

This detailed, real-time data is valuable for public health applications. For example, it has been proven effective in monitoring compliance with measures like social distancing, tracking quarantine adherence during the pandemic, and identifying potential outbreak hotspots \citep{Alo2022}. However, the use of such personal data necessitates robust privacy protections. Standard practices include anonymization, aggregation, and clear user consent managed through device settings and user agreements. To further enhance privacy, advanced techniques like federated learning can be employed, allowing analytical models to be trained on decentralized data without exposing raw location information. By leveraging these ethically managed IoT insights, public health authorities and researchers can complement other data sources, more effectively track disease spread, and implement timely interventions.

\begin{figure*}[htbp]
\centering
\includegraphics[width=0.9\textwidth]{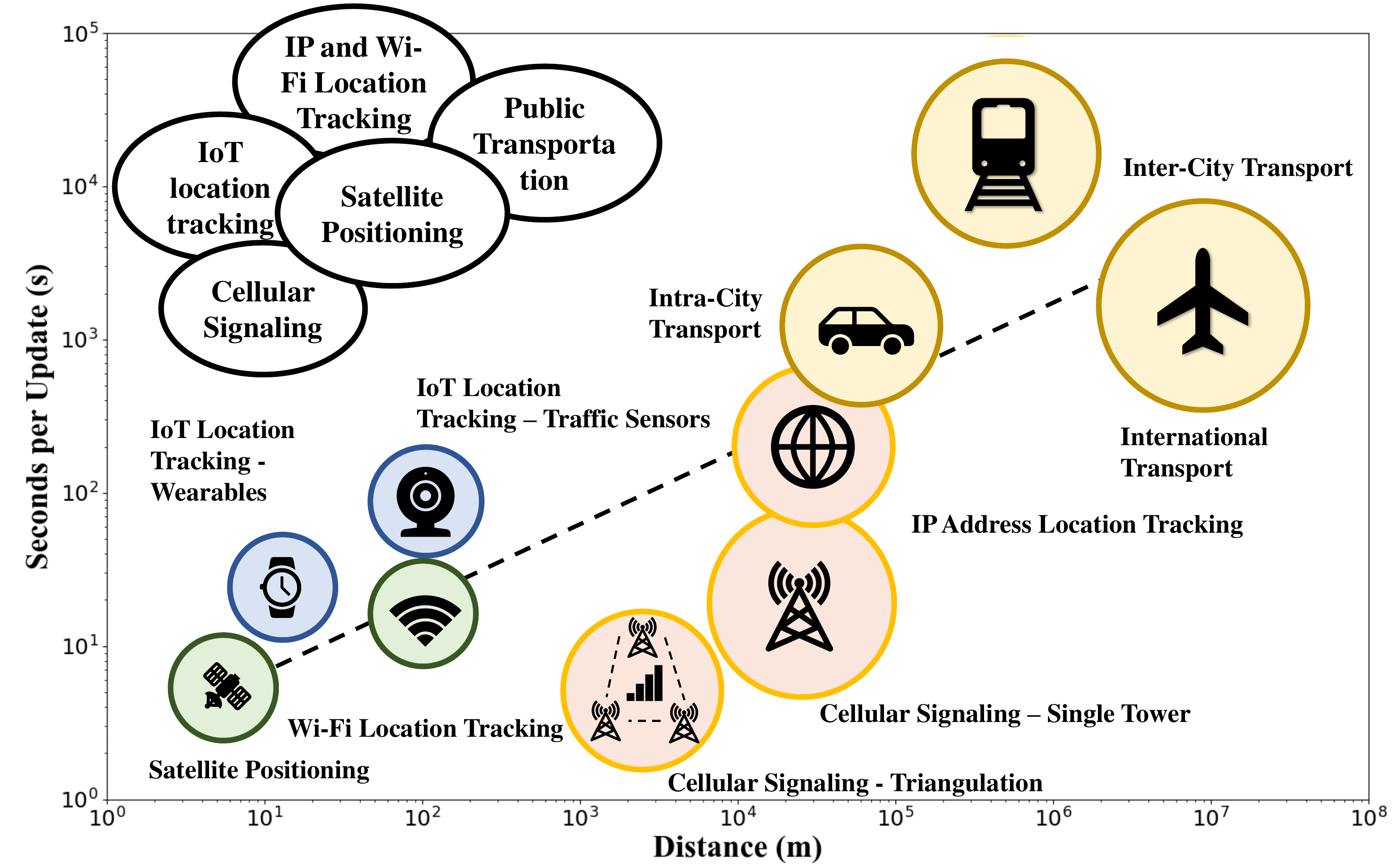}
\caption{Estimated spatiotemporal resolution of human mobility data sources.}

{\small\noindent
    \parbox{\textwidth}{
The radii of the circles represent data granularity, while the X-axis and Y-axis indicate the spatial distance of human movement and the temporal interval between record updates, respectively. It is important to note that the boundaries between different human mobility data sources are often ambiguous. For instance, Location-Based Service (LBS) data from platforms such as Foursquare, Twitter, Facebook, WeChat, and Weibo can fall under multiple categories, including IP-based, Wi-Fi-based, and cellular signaling. These platforms typically determine user locations using IP addresses and Wi-Fi networks, while also leveraging cellular signals to enhance positioning accuracy—particularly in scenarios where Wi-Fi is unavailable or unreliable.
}}

\label{Fig.3}
\end{figure*}

It is worth noting that a number of challenges may compromise the accuracy and ethical application of epidemic models. First, data biases, such as uneven geographic coverage, underrepresentation of rural or marginalized populations, and demographic skews, can lead to inequitable or misleading modeling outcomes \citep{Santana2023}. Privacy concerns are also paramount, particularly with fine-grained data from GPS, cellular signaling, and IoT sources, where even anonymized data may still allow for re-identification. Moreover, the boundaries between data types are often blurred \citep{Xu2022}; Location-Based Service (LBS) data from platforms like Facebook and WeChat may fall simultaneously under GPS, IP, and cellular categories, making categorization and source attribution difficult \citep{Barreras2024}. In terms of interoperability, varying data formats, update frequencies, and spatial-temporal resolutions across sources complicate direct comparisons and integration efforts. \autoref{Fig.3} illustrates the measurement ranges and update intervals for various sources of human mobility data. The radius of each circle represents the level of data granularity, with larger circles indicating more aggregated mobility data points. Public transportation data may be collected via Bluetooth or CRFID, while IoT and GPS data typically transmit over Wi-Fi or cellular networks, further fragmenting the data ecosystem \citep{Hernandez-Orallo2020}. Overcoming these challenges requires the development of standardized protocols, robust ethical frameworks, and advanced data fusion methods to ensure responsible and effective use of mobility data in epidemic response.

\section{Representation of Human Mobility}\label{sec:3}
\subsection{Trajectory}\label{sec:3.1}
A trajectory refers to the movement path of an individual over time, typically represented as a sequence of geographic locations. Trajectory data is the primary data form of human mobility, which can (technically speaking) be captured from various sources such as social media platforms (Facebook, Twitter), GPS devices, and navigation applications (Google Maps, Baidu Map, and Amap). It is widely used in studies of urban planning, travel behavior, and emergency response \citep{Li2020a,Zhang2022a}.

\begin{figure*}[htbp]
\centering
\includegraphics[width=0.6\textwidth]{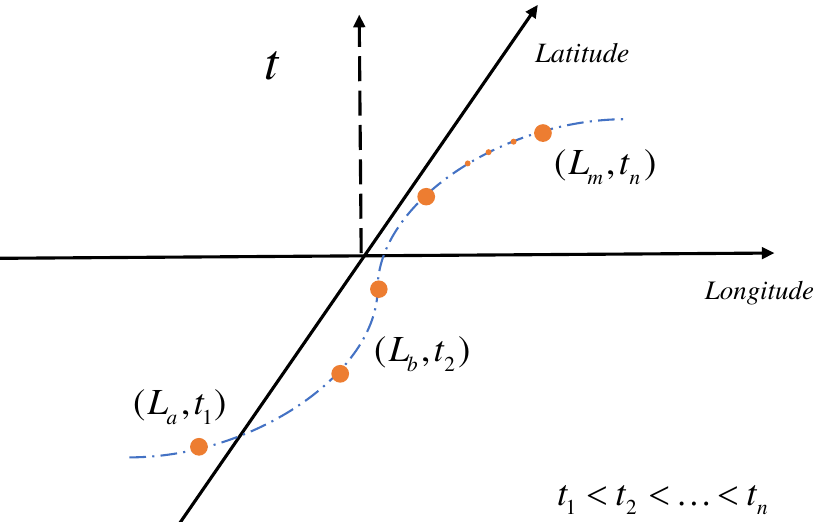}
\caption{Trajectory data example.}

{\small\noindent
    \parbox{\textwidth}{
The horizontal and vertical axes correspond to longitude and latitude, respectively, while the dash line illustrates the progression of time along the trajectory. The curve between sampled points visualizes the inferred path connecting consecutive observations, describing the alternation between stay (stationary) and displacement (moving) states that define a trajectory.
    }
}

\label{Fig.4}
\end{figure*}

Generally, a trajectory can be seen as an alternation of stay and displacement (e.g., two states in which the individual spends time at a definite location and in which the individual moves between locations). It can be represented as a connotative chronological sequence ${L_1} \to {L_2} \to ... \to {L_n}$ , where   denotes the  th location, and the x-axis and y-axis represent longitude and latitude, respectively, as shown in \autoref{Fig.4}. Trajectory data can be used to build detailed co-location or contact networks (see \autoref{sec:5.3.2}), identifying times and locations where individuals are likely to interact, which are essential factors in infectious disease modeling \citep{Alo2022,Machens2013}.

Trajectory data forms the basis for digital contact tracing during pandemic outbreaks by enabling the identification of close contacts and potential exposure events. This data allows health authorities to trace the movements of infected individuals and determine where and when they may have interacted with others, identifying potential transmission chains. However, the excessive granularity of trajectory data raises significant privacy concerns, as it can reveal sensitive information about individuals' daily routines, locations visited, and personal habits \citep{Lu2013}. Consequently, after the COVID-19 pandemic, most trajectory datasets like Apple Mobility Trends Reports are no longer collected and publicly available for researchers.

\subsection{Mobility Network}\label{sec:3.2}
\begin{figure*}[htbp]
\centering
\includegraphics[width=\textwidth]{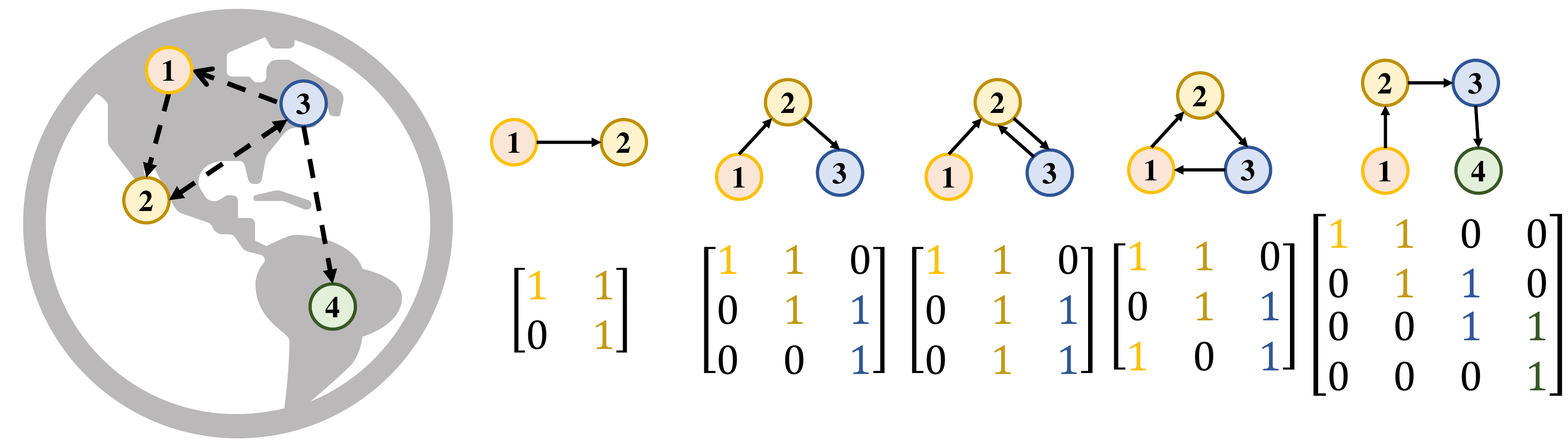}
\caption{Example of the individual mobility network. }

{\small\noindent
    \parbox{\textwidth}{
Each node is labeled with a number indicating the order of visits, and each edge is directed, indicating the origin-destination of trips made between the connected locations.
    }
}
\label{Fig.5}
\end{figure*}

In principle, human mobility is the combined information of where people are as a function of time, but depending on data availability, questions asked about the data, and the disease investigated, one would try to assemble a more coarse-grained representation (e.g., mobility network) rather than the raw trajectories. A mobility network often refers to a graph-based representation of how individual or populations move between geographic locations, where nodes correspond to locations and edges represent the movements between these locations, as shown in \autoref{Fig.5}.

One of the most common forms of mobility networks is the Origin-Destination (OD) matrix, which captures the number of movements between different locations, defined as origins and destinations, over specific time intervals. In an OD matrix ${{\bf{A}}_{{\bf{OD}}}}$, where $a_{ij}$ and $a_{ji}$ are the outflow and inflow from location $i$ to $j$, respectively. The temporal aspect of the OD matrix allows researchers to capture dynamic changes in population movement, enabling analysis of daily commuting cycles, weekend variations, seasonal trends, and mobility responses to epidemic interventions. The backflow model provides an alternative representation of a mobility network. For each time period $k$, the unidirectional net flow matrix can be denoted as $E^k$, whose elements of $E^k$ are calculated by ${e_{ij}} = {a_{ij}} - {a_{ji}}$ (${e_{ji}} = {a_{ji}} - {a_{ij}}$). This backflow model can be used to describe the net inflow and outflow mobility pattern in pandemics \citep{Tan2021a}. The OD matrix and its variants are standard analytical tools for examining population flow, capturing the movement behaviors and patterns of large groups across different regions \citep{Bengtsson2015,Tan2021a}. In the context of a pandemic, these tools help describe and quantify population movement and settlement patterns, which is crucial for controlling virus transmission, optimizing public health strategies, and designing more effective preventive measures (see \autoref{sec:7}) \citep{Xian2025}.

\subsection{Contact Network}\label{sec:3.3}
A contact is, in general, a pair of people and a time interval when they are in sufficient proximity for a disease to spread \citep{Masuda2017}. What a contact is depends strongly on the pathogen; for some diseases, people would not even have to be within visual range for a contact to happen \citep{Liljeros2007}. In temporal network epidemiology, time is usually discretized for measurement technical reasons, and a contact is thus a triple $(i,j,t)$ stating that individuals   and   have been in contact at time (interval) $t$ \citep{Holme2016,Du2020b}.

Mathematically, a contact network is also a graph representation of individuals (nodes) and their interactions or contacts (edges) that can lead to the spread of epidemics \citep{Leung2021,Mossong2008,Starnini2013}. It can be represented as a contact matrix $C$, and the element $c_{ij}$ represents the contact between individuals $i$ and $j$. Specifically:
\begin{equation}
{\bf{C}} = \left( {\begin{array}{*{20}{c}}
{{c_{11}}}&{{c_{12}}}& \cdots &{{c_{1n}}}\\
{{c_{21}}}&{{c_{22}}}& \cdots &{{c_{2n}}}\\
 \vdots & \vdots & \ddots & \vdots \\
{{c_{n1}}}&{{c_{n2}}}& \cdots &{{c_{nn}}}
\end{array}} \right)\text{.}
\label{eq:1}
\end{equation}
If $c_{ij}$ is greater than $0$, it indicates that there is a contact between individual $i$ and $j$, with the value represents the frequency or strength of the contact. If $c_{ij}$ is $0$, it suggests no contact between individual $i$ and $j$. By illustrating who interacts with whom and how frequently these interactions occur, contact networks enable epidemiologists to simulate the dynamics of disease transmission accurately, predict potential outbreak patterns, and identify critical intervention points. However, many real-life interactions occur within larger social contexts, such as households, schools, workplaces, or community gatherings, where multiple individuals simultaneously interact in groups rather than in simple pairwise encounters . These group interactions create clusters within the network, significantly affecting the overall connectivity and the speed at which infections propagate (see \autoref{sec:4.2}) \citep{Bengtsson2014,Salathe2010}. 

Unlike mobility networks, which emphasize aggregated flows between locations, contact networks focus on connections at the individual level. This allows for more granular representation of human movement and interaction, making them particularly relevant for studying contact-based infectious diseases \citep{Liu2018,Machens2013}. A landmark study conducted between May 2005 and September 2006 across eight European countries (Belgium, Germany, Finland, Great Britain, Italy, Luxembourg, Netherlands, and Poland) systematically quantified how social contacts and mixing patterns impact the spread of infectious diseases across different populations \citep{Mossong2008}. This pioneering work represented the first large-scale quantitative survey of contact patterns relevant to infection transmission in Europe, laying the empirical foundation for constructing age-structured contact matrices widely used in epidemiological modeling.

\subsection{Mobility Trend and Index}\label{sec:3.4}
Relying solely on a data type makes it challenging to derive comprehensive and realistic human mobility patterns during epidemics. For instance, individual mobility trajectories and network data alone cannot capture actual contact events between susceptible individuals. Conversely, relying solely on contact network data fails to provide a comprehensive view of movement trajectories and cross-regional population flows. Furthermore, interdisciplinary studies often integrate various non-mobility datasets (e.g., demographic, transportation, social media, human behaviors, and geographic data) into human mobility datasets to enhance the analysis and understanding of epidemics \citep{Chande2020}. However, from the perspective of data providers, who frequently possess additional information that could aid in understanding human mobility patterns, it is typically not feasible to share such data publicly due to concerns over business competition, legal and regulatory constraints, and privacy issues. As a result, many studies that incorporate population mobility into epidemic modeling rely on publicly availab\citep{Guo2021,Qi2023}le mobility trends and indices, which are typically composites of diverse data types and sources.

Mobility trend and index datasets involve the aggregation and transformation of multi-source heterogeneous anonymous spatiotemporal mobility data, which presents processed data to the public to ensure privacy protection while providing insights. Prominent examples of such practices include Google Community Mobility \citep{Guo2021,Qi2023}, Apple Mobility Trends \citep{Strzelecki2022}, Spectus Mobility Data \citep{2024q}, Baidu Migration Index \citep{Wei2020}, Descartes Lab Mobility Index \citep{2024a}, and Facebook Data for Good \citep{2025w}. \autoref{tab.2} summarizes the profiles of selected human mobility datasets. They enable researchers to make more accurate predictions and better-informed decisions by leveraging the strengths of each data type while mitigating their respective limitations.

\begin{landscape}
\begin{table}[htbp]
    \fontsize{8pt}{10pt}\selectfont  
    \centering
    \caption{Profile of selected human mobility datasets.}
    \label{tab.2}
    
    \begin{tabular}{@{}p{1.8cm} p{1.8cm} p{2.5cm} p{1.8cm} p{1.5cm} p{1.8cm} p{1.8cm} p{5cm} p{3cm}@{}}

        \toprule
        \textbf{Data} & \textbf{Mobility Source} & \textbf{Mobility Type} & \textbf{Provider} & \textbf{Region} & \textbf{Available Time} & \textbf{Publicly Available} & \textbf{Descriptions} & \textbf{Accessibility} \\
        \midrule

        Google Community Mobility & 
        GPS/Wi-Fi/IoT/IP & 
        Mobility trend and index & 
        Google & 
        Global & 
        2020/2/15 - 2022/10/15 & 
        Yes &
         \textbullet The dataset shows the changes in visits and length of stay at different locations.
        
         \textbullet Data comes from users who have opted-in to Location History for their Google Account. & 
        \url{https://www.google.com/covid19/mobility} \\
    \addlinespace
        Apple Mobility Trends & 
        GPS/Wi-Fi/IOT/IP & 
        Mobility trend and index & 
        Apple & 
        Global & 
        2020/1/13 - 2022/4/12 & 
        Yes & 
         \textbullet The dataset includes daily changes in requests for directions on the Maps app by driving, transit, and walking for several spatial levels.
        
         \textbullet Data is collected via Apple Maps using anonymized location services. & \url{https://www.kaggle.com/datasets/caseycushing/apple-mobility} \\
    \addlinespace
        Baidu Migration Index & 
        GPS/Wi-Fi/IP & 
        Mobility network & 
        Baidu & 
        China Mainland & 
        Up to now & 
        No & 
         \textbullet The dataset shows the number of people migrating from one city to another city.

         \textbullet Data comes from (1) the users of Baidu Map, (2) Third-party Apps, and (3) government data. & 
        \url{https://qianxi.baidu.com} \\
    \addlinespace
        DL-COVID-19 Mobility Statistics & 
        GPS & 
        Mobility trend and index & 
        Descartes Labs & 
        America & 
        2020/3/1 - 2021/4/20 & 
        Yes & 
         \textbullet The mobility statistics (representing the distance a typical member of a given population moves in a day) at the U.S. state and county level.

         \textbullet The datasets come from commercially available mobile device location datasets using cloud computing resources. & 
        \url{https://github.com/descarteslabs/DL-COVID-19} \\
    \addlinespace
        Spectus Mobility Data & 
        GPS/Wi-Fi/IP & 
        Trajectory/Mobility network & 
        Cuebiq & 
        America & 
        Up to now & 
        No & 
        \textbullet Spectus provides anonymous mobility data via a platform-as-a-service (PaaS).

        \textbullet Spectus mobility data is collected from its partner smartphone applications, whose location is at the core of the app's functionality. & 
        \url{https://spectus.ai} \\
    \addlinespace
        Data for Good & 
        GPS/Wi-Fi/IP & 
        Mobility trend and index & 
        Facebook (Meta) & 
        Global (except specific regions) & 
        Up to now & 
        Yes & 
         \textbullet The data comes from Facebook users who opt into location history and background location collection.
         
         \textbullet The amount of movement is quantified by counting the number of 600m x 600m areas a person is observed in within a day.
         & 
        \url{https://github.com/BDBC-KG-NLP/COVID-19-tracker} \\
        
        \bottomrule
    \end{tabular}%
\end{table}
\end{landscape}

    

\section{Interplay Between Human Mobility, Contact and Epidemiology}\label{sec:4}
In this section, we introduce studies that reveal the correlations between mobility, contact behavior, and disease spread, and further explore how different forms of human movement contribute to transmission dynamics across spatial, temporal, and social scales. 

\subsection{Transmission by Population Flow}\label{sec:4.1}
Transmission by population flow in epidemics refers to the spread of infectious diseases facilitated by the movement of people across geographic regions \citep{Liu2024d,Wei2020,Xiong2020,Belman2024}. This mechanism encompasses following dimensions: (1) the role of long-distance travel by infected individuals in disseminating pathogens across regions; (2) the significance of transportation hubs, such as airports, in early-stage epidemic transmission; (3) the correlation between mobility intensity and both local and international disease spread; (4) the impact of seasonal and demographic mobility fluctuations on transmission dynamics; and (5) the need for targeted interventions in high-traffic areas and among mobile, vulnerable groups.

\begin{figure*}[htbp]
\centering
\includegraphics[width=\textwidth]{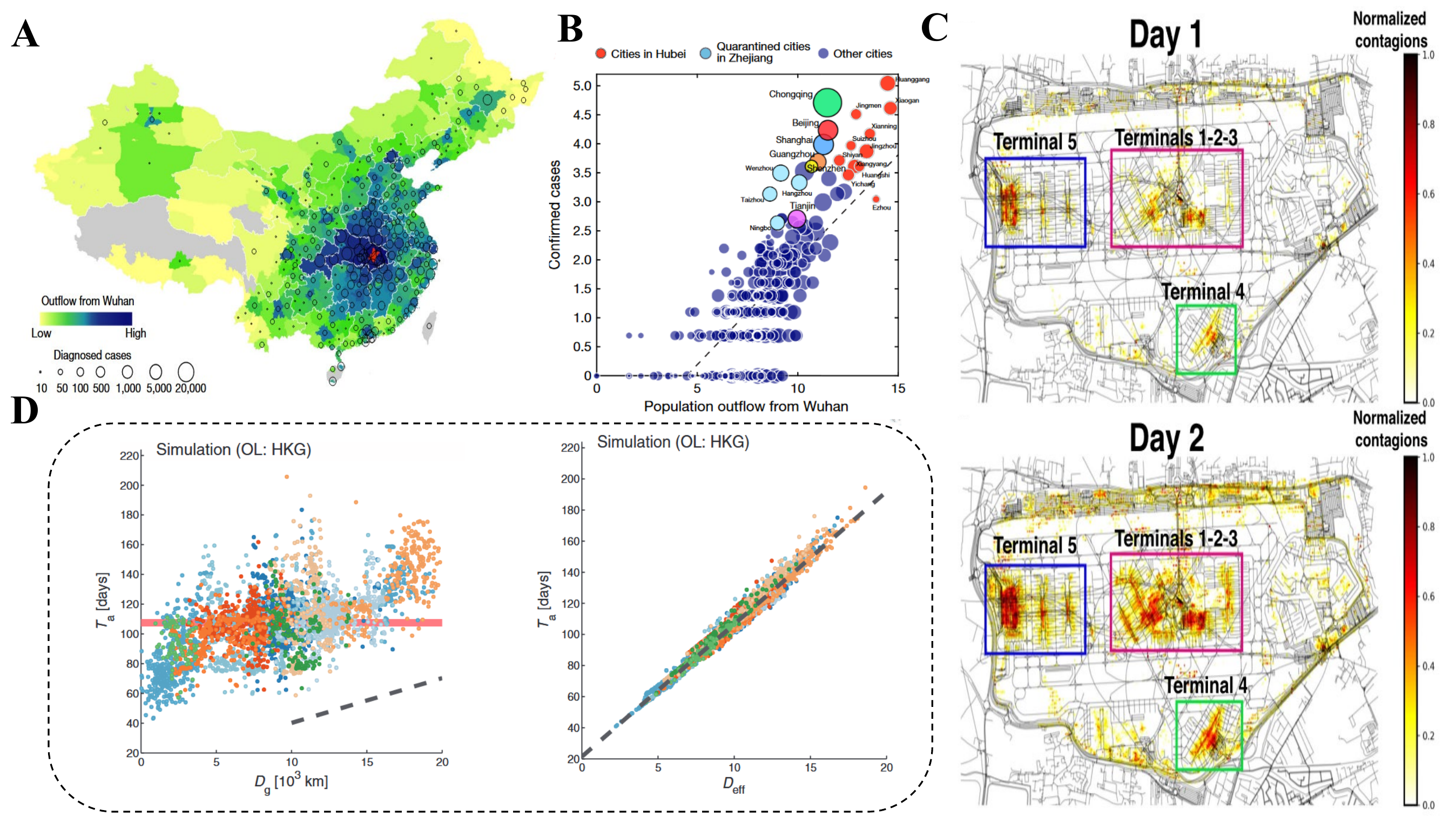}
\caption{Correlation between population flow and epidemic transmission.}

{\small\noindent
    \parbox{\textwidth}{
(A) Heat map of outflow from Wuhan during the early stage of COVID-19, highlighting major destinations across mainland China. There is a high overlap between the geographical distribution of aggregate population outflow from Wuhan until 24 January 2020 (in red) and the number of confirmed cases of COVID-19 in other Chinese prefectures (n = 296 prefectures). (B) The  relationship between the log-transformed aggregate population outflow from Wuhan (up to 24 January 2020) and the log-transformed number of confirmed cases by prefecture on 26 January 2020. (C) Heatmap of the SIR model simulation within an airport environment, showing the cells in the airport area where infections occur. At the top after one simulation day and at the bottom after two days. The color scale is normalized for each period considered. (D) Epidemic arrival time plotted against geographic distance from the source, and against effective distance derived from the mobility network. The left panel illustrates the relationship between the epidemic arrival time ($T_a$) and the geographic distance ($D_g$) from the outbreak location (OL) for all 4,069 nodes in the global mobility network. Each point represents a location. Although the arrival time generally increases with distance, the correlation between ${T_a}$ and ${D_g}$ is weak (${R^2}{\rm{ = }}0.34$), indicating substantial variability in arrival times among locations at similar distances. The right panel presents the relationship between the epidemic arrival time (${T_a}$) and the effective (${D_{eff}}$), and the arrival time shows a strong linear relationship with effective distance (${R^2}{\rm{ = }}0.973$). This finding indicates that effective distance derived from mobility flow probabilities rather than physical distance serves as an excellent predictor of epidemic arrival times. Fig. 6 (A) and Fig. 6 (B) are from Fig. 1 (a) and Fig. 2 (a) in Ref. \citep{Jia2020}, while Fig. 6 (C) and Fig. 6 (D) are from Fig. 2 in Ref. \citep{Mazzoli2023} and Fig. 1/2 (C) in Ref. \citep{Brockmann2013}, respectively.
    }
}

\label{Fig.6}
\end{figure*}

Human mobility exhibits scale-free characteristics, whether long-distance travel or local commuting \citep{Alessandretti2020,Barabasi1999,Graf2024,Tang2023,Yang2023}, which means that a small number of individuals with unusually high mobility can disproportionately contribute to pathogen spread \citep{Iannelli2017,Zhang2023a}. For example, infected individuals with frequent and long-distance travel have the potential to expand a localized outbreak into a regional or global health emergency fast. The examination of the correlation facilitates a comprehensive understanding of, and the ability to anticipate, transmission patterns and implement timely control strategies \citep{Marguta2015}. One clear example of this correlation is the early-stage predicting role of transportation hubs in global disease outbreaks. Generally, international flight networks link otherwise distant regions, enabling pathogens to travel thousands of miles in mere hours. Similarly, densely interconnected urban public transit systems, especially those involving enclosed and crowded spaces, facilitate the rapid local spread of airborne and contact-transmissible diseases, like influenza or SARS-CoV-2 \citep{Mitze2022,Pepe2020,Tan2021a}. Previous studies show that domestic transmission within a country can show a strong correlation between the population outflow and the number of cases reported, as depicted in \autoref{Fig.6} (A, B) \citep{Jia2020}. During the initial phase of COVID-19 outbreak, Chinese cities receiving large numbers of travelers from Wuhan (the place where the first confirmed case found) exhibited significantly higher case counts \citep{Lai2022a}. This association underscores the importance of focusing control measures on critical points of connectivity, such as railway stations, airports, and customs ports, where a large number of individuals interact and disperse.

The role of local contacts is equally important. \autoref{Fig.6} (C) illustrates that the infection can initially disseminate from an index case to nearby individual and subsequently propagate through the spatial layout of the facility as individuals traverse the airport \citep{Mazzoli2023}. Such hubs function as microcosms of epidemic processes, where short-duration, high-density contacts accumulate into macro-level transmission beyond transportation environments \citep{Tan2021a}. Beyond these physical spaces, the concept of effective distance, a metric derived from network topology rather than geographic distance, reliably predicts disease arrival times, as shown in \autoref{Fig.6} (D). Even if underlying epidemiological parameters remain unknown, mobility-based effective distance explains disease spread more accurately than linear geographic proximity, enabling early detection of transmission routes for outbreaks such as the worldwide 2009 H1N1 influenza pandemic and 2003 SARS epidemic \citep{Brockmann2013}.

Across various spatial scales, studies have shown that population flow is a more accurate predictor of infection spread than relatively static variables such as population size, economic status, or geographic distance from the epidemic origin \citep{Jia2020,Rader2020}. This is primarily because most infectious diseases require physical contact or proximity, both of which are facilitated by human movement. This relationship has been particularly evident in past pandemics, where the speed, extent, and intensity of transmission have closely mirrored mobility patterns \citep{Tizzoni2012,Zhang2020a}. Some researches further suggest that specific types of movement such as frequent commuting between densely populated urban centers are especially conducive to accelerated transmission \citep{Schlapfer2021a}. Additionally, variations in contact rates and mobility behaviors across different populations and regions introduce further heterogeneity into disease dynamics and detection efforts \citep{Dalziel2013}.

Transmission risk likely increase substantially during periods of seasonal or demographic mobility surges \citep{Charu2017,Petrova2018,2025i}. Holidays, school breaks, and labor migration events are marked by elevated social contact and crowding, creating ideal conditions for disease amplification during times of high population flux \citep{Guo2024}. Vulnerable demographic groups, such as migrant workers, face higher exposure risk due to their frequent movement and often substandard living and working conditions in areas where these populations converge, such as urban slums and temporary housing camps with limited access to healthcare \citep{Du2022b,Lai2022,Murray2021,Pullano2020}. For instance, the resurgence of Ebola virus in Guinea in 2021 has been linked to population movement from affected areas \citep{Keita2021}. Genetic analysis showed that the viral genomes closely clustered with those from the earlier outbreak, suggesting that the new outbreak did not result from a novel spillover event from an animal reservoir, but rather from human-to-human transmission.

In addition, some migrant groups may originate from countries or regions with limited or disrupted healthcare and vaccination systems, leading to misalignment with the vaccination schedules of their host areas. This immunization gaps increase these populations' susceptibility to infection and may amplify outbreak risks in host populations \citep{Deal2021}. Such findings highlight the need for targeted health interventions and policies to address mobility and immunization gaps in migrant populations \citep{Chen2020}. It is significant to quantify the impact of population movement from one location to another and assess its influence on infectious disease transmission. Numerous studies suggest that human population movements are not random but exhibit high regularity and predictability across various spatial and temporal scales. These become a crucial factor in evaluating the effect of epidemic spread \citep{Barabasi2005}. \autoref{tab.3} highlights six typical human mobility models used to quantify this impact. Each model defines the flow $T_{i,j}$ between region $i$ and $j$, and provides unique insights into how human mobility contributes to disease spread. The models differ in how they account for population, distance, and opportunities, which play a specific role in simulating epidemic dynamics.

\begin{landscape}
\begin{table}[htbp]
    \fontsize{8pt}{10pt}\selectfont  
    \centering
    \caption{Overview of population flow models in epidemic modeling.}
    \label{tab.3}

    \begin{tabular}{@{}p{3cm} p{4cm} p{6cm} p{10cm}@{}}
        \toprule
        \textbf{Model} & \textbf{Formula} & \textbf{Explanation} & \textbf{Role in Epidemic modeling} \\
        \midrule

        Gravity Model & 
        ${T_{ij}} = \frac{{G \cdot {P_i}^\alpha  \cdot {P_j}^\beta }}{{d_{ij}^\gamma }}$ & 
        $P_{i}^\alpha$ and $P_{j}^\beta$ are the populations; $\alpha$ and $\beta$ control the influence of population sizes, normally are $1$; $d_{i j}$  is the distance between areas; $\gamma$ controls the deterrence effect of distance; $G$ is a constant.& 
        Predicting how infectious diseases are transmitted between regions based on population size and distance. \\
    \addlinespace
        Impedance Model & 
       ${T_{ij}} = \frac{{{P_i} \cdot {P_j}}}{{d_{ij}^\gamma {\rm{ + }}\varepsilon }}$ & 
        $P_{i}$ and $P_{j}$ are the populations; $d_{i j}$ is the distance or resistance factor; $\gamma$ is the impedance factor that controls the effect of distance; $\varepsilon$ is the resistance term. & 
        Accounting for barriers or resistance (e.g., travel cost, time, distance) that may slow down disease spread. It helps model how travel restrictions or lockdowns can reduce transmission rates.\\
    \addlinespace
        Intervention Opportunity Model & 
        ${T_{ij}} = \frac{{{O_i} \cdot {D_j}}}{{\sum\limits_{k \le j} {{O_k}} }}$ & 
        $O_i$ is the opportunity at origin $i$, $D_j$ is the attraction of destination $j$; $k \le j$ indicates the cumulative opportunities encountered before reaching destination $j$, including $j$ itself.& 
        It captures the spread of disease by considering the final destination and the intermediate stops where individuals might be exposed to the disease. Understanding how travel hubs like airports contribute to disease transmission is essential. \\
    \addlinespace
        Radiation Model & 
        ${T_{ij}} = \frac{{{P_i} \cdot {P_j}}}{{({P_i} + {s_{ij}})({P_i} + {P_j} + {s_{ij}})}}$ & 
        $P_i$ and $P_j$ are the populations; $s_{ij}$ is the number of opportunities between the two areas. & 
        Modeling disease spread based on opportunity distribution, not distance. It helps to simulate the regional spread of diseases without relying on arbitrary parameters like distance, particularly for diseases that do not rely heavily on distance for transmission.\\
    \addlinespace
        Population Weighted Opportunities Model & 
        $T_{i j}=P_{i} \cdot \frac{O_{j}}{\sum_{k} O_{k} \cdot W_{k}}$ & 
        $O_i$ is the opportunity at origin $j$, $W_k$ is the population weight in other areas. & 
        Incorporating population density into disease modeling to predict how densely populated areas are more likely to experience faster and broader spread of infections.\\
    \addlinespace
        Opportunity Priority Selection Model & 
        $T_{i j}=\frac{O_{j}^{\alpha}}{\left(d_{i j}+1\right)^{\beta}}$ & 
        $O_i$ is the opportunity at origin $i$, $d_{ij}$ is the distance between areas $i$ and $j$; $\alpha$ and $\beta$ are adjustable parameters for opportunity and distance. & 
        Emphasizing the importance of specific destinations based on their priority (e.g., work or education opportunities), which helps identify high-risk regions for disease outbreaks. \\

        \bottomrule
    \end{tabular}%
\end{table}
\end{landscape}

\subsection{Spread through Contact Behaviors}\label{sec:4.2}
Contact behaviors refer to the patterns by which people interact with each other, including how often, how long, and in what situations those interaction occur \citep{Crawford2022,Zhang2020a}. These patterns shape the structural foundation for transmission and are critical for understanding the dynamics of infectious disease spread, informing both epidemic modeling and the design of intervention strategies \citep{Mossong2008,Schnyder2025}. The interplay between disease progression and human behavior is bidirectional: as infections spread, they often alter social behavior, which in turn reshapes future transmission risk. Comprehending this feedback loop is crucial for building realistic models and developing interventions that not only mitigate disease spread but also consider the societal costs and behavioral responses \citep{Aleta2022}. By examining how contact frequency, duration, and social contexts vary across different groups, we can enhance the granularity and accuracy of predictive epidemic models for tailoring targeted interventions \citep{Eubank2004}. Research on epidemic spread through contact behaviors focuses on three main aspects: (1) frequency and duration of contact, (2) social context and environmental factors, and (3) demographic and behavioral heterogeneity. 

\begin{figure*}[htbp]
\centering
\includegraphics[width=\textwidth]{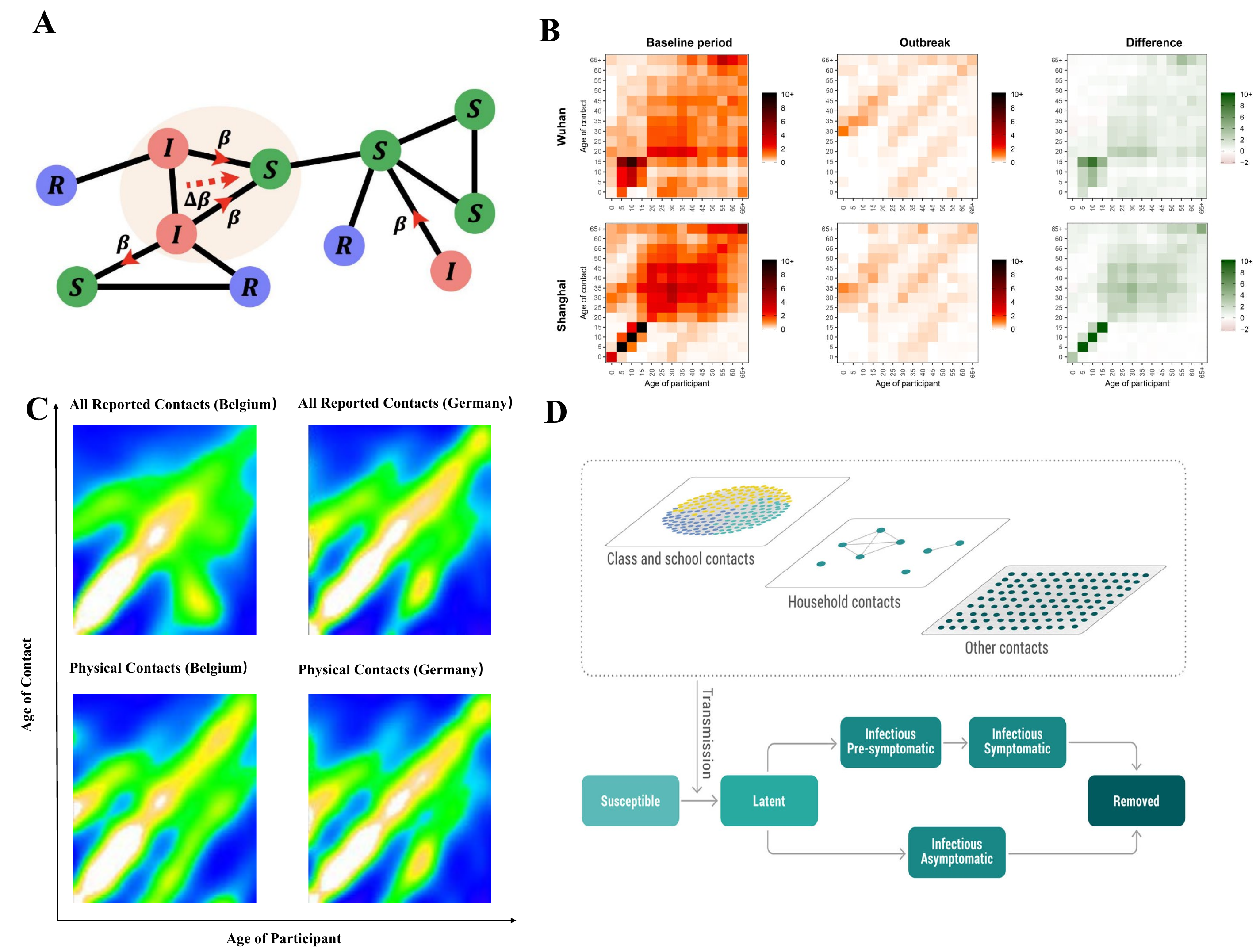}
\caption{Epidemics spreading through contact patterns.}

{\small\noindent
    \parbox{\textwidth}{
 (A) Transmission of an infectious agent across a contact network, illustrating how person-to-person links facilitate the spread of disease. (B) Age-stratified contact matrices from empirical contact data, showing higher interaction rates within similar age groups. The contact matrix illustrates the average number of daily contacts between individuals of different age groups in Wuhan and Shanghai on regular weekdays. Each cell in the matrix indicates the mean number of contacts that individuals in one age group have with those in another, with color intensity reflecting contact frequency. The matrices were constructed using bootstrap sampling with replacement, weighted by the actual population age distribution in each city, and averaged over 100 bootstrapped realizations. (C) Smoothed contact matrices for Belgium and Germany. All reported contacts and physical contacts are weighted by sampling weights  white indicates high contact rates, green intermediate contact rates, and blue low contact rates, relative to the country-specific contact intensity. Fitting is based on a tensor-product spline to contact matrix data using a negative binomial distribution to account for overdispersion. (D) Schematic representation of the transmission model and contact networks in schools, households, and other places. Each environment features distinct connectivity patterns that shape transmission dynamics. Fig. 7 (A), Fig. 7(B), Fig. 7(C) and Fig. 7 (D) are from Fig. 3 in Ref. \citep{Chen2024c}, Fig. 1 in Ref. \citep{Zhang2020a}, Fig. 3 in Ref. \citep{Mossong2008} and Fig. 5 in Ref. \citep{Liu2022c}, respectively.
    }
}

\label{Fig.7}
\end{figure*}

The frequency and duration of inter-host contacts are fundamental determinants of transmission probability \citep{Zhang2019}. High-frequency or prolonged contacts are more likely to lead to infections, particularly for pathogens transmitted via droplets, aerosols, or physical contact. Understanding the structures and dynamics of contact networks informs parameter estimation, interpretation, and the formulation of control measures such as social distancing, mask mandates, or quarantine guidelines\citep{Fenichel2011,Harish2024}. \autoref{Fig.7} (A) illustrates an example of disease propagation across a contact network, showing how an infectious individual transmits the pathogen to connected contacts over time \citep{Chen2024c}. Adaptive changes in behavior (e.g., by avoiding crowded places, wearing masks, or adhering to quarantine guidelines) can reshape the contact network, impact infection rates, and reduce epidemic spread. These behavioral shifts imply that traditional static models of disease transmission, which assume fixed contact rates, may not catch fully the complexity of real-world epidemic dynamics. Understanding the relationship among human mobility, contact patterns, and disease transmission involves constructing appropriate contact networks and applying network analytics within models \citep{Ferretti2024,Huang2021,Kendall2024,Selinger2021,2022c}.

The social context and environmental factors in which interactions occur greatly influences transmission risk. Contact intensity and structure differ across settings, such as households, schools, workplaces and transportation hubs, as well as between groups with different social roles. Individuals with disparate social identities exhibit disparate contact patterns at a more localized level, giving rise to divergent transmission dynamics \citep{Brockmann2013,Pappalardo2015,Tatem2006}. For instance, at London Heathrow Airport, airport staffs had denser contact networks than transient passengers, making an infected worker far more likely to propagate disease than a single infected passenger \citep{Mazzoli2023}. In many respiratory disease outbreaks, interventions like school closures can potentially reduce peak incidence by up to 40-60\% and delay the epidemic peaks. However, such interventions are often insufficient alone and must be deployed with broader strategies. Incorporating these social and environmental layers into epidemic models allows us to capture the nuanced, real-world transmission dynamics shaped, enabling better precision in identifying high-risk interactions and optimizing targeted interventions \citep{Wesolowski2017}.

Demographic traits (e.g., age, gender, and occupation) also shape contact patterns and, consequently, disease dynamics. Some infectious diseases, such as HIV, requiring prolonged or intimate contact for transmission, necessitate a more profound comprehension of the behavioral and social factors that contribute to their transmission dynamics \citep{Liljeros2001}. In such case, the focus of prevention efforts must shift to understanding and addressing the exposure patterns within high-risk groups \citep{Okello2024}. These include the frequency and nature of interactions within specific contact networks, such as those involving sexual partners or intravenous drug users. Identifying super-spreaders or hotspots of exposure within these networks can be crucial for implementing targeted interventions \citep{Nardell2024}. Besides, age-specific contact patterns offer another layer of insight. Empirical studies demonstrate that people are most likely to interact, and thus infect others within their own age group, especially in environments like schools and households \citep{Mossong2008,Zhang2020a}. This age-assortative mixing is visualized in \autoref{Fig.7} (B, C), where contact matrices show stronger interactions within specific age bands. \autoref{Fig.7} (D) further demonstrates the schematic representations of these contact and transmission networks across different circumstances, such as schools, households, and other social settings \citep{Liu2022c}.

Analyzing contact patterns across these three dimensions (temporal dynamics, contextual settings, and demographic profiles) enables targeted and effective modeling and public health response \citep{Wang2021,Zhang2024e}. In particular, contact tracing conducted by field epidemiological investigations has proven instrumental in identifying and isolating cases within tightly connected social clusters \citep{Hossain2022,Pang2020}. In contexts where close-contact transmission dominates, tracing the structure of social interactions becomes critical to breaking chains of infection. Besides, the utilization of digital contact tracing via mobile phone signaling, sensors and Bluetooth-based proximity data can now track interactions in real-time, allowing for rapid outbreak detection and early warning systems \citep{Ferretti2024,Kendall2024,Rodriguez2021}. This advancement allows for the quicker identification of transmission chains, enhancing our ability to respond to outbreaks with greater speed and precision \citep{Firth2020,Lai2020,Zhang2020a}.

\section{Epidemics Models Incorporated Human Mobility}\label{sec:5}

\subsection{Compartmental Models}\label{sec:5.1}
Compartmental models represent a foundational class of population-based epidemiological modeling frameworks that simulate disease transmission dynamics by dividing populations into discrete compartments corresponding to different stages of infection, such as Susceptible (S), Exposed (E), Infectious (I), and Recovered (R) \citep{Castillo-Chavez2016,Li2021d,Liu2023d,Sallah2017,Tan2021,Urabe2016,Wang2017,Zhang2023e}. These models typically rely on the assumption of homogeneous mixing, meaning that each individual has an equal probability of coming into contact with others. 

\begin{figure*}[htbp]
\centering
\includegraphics[width=\textwidth]{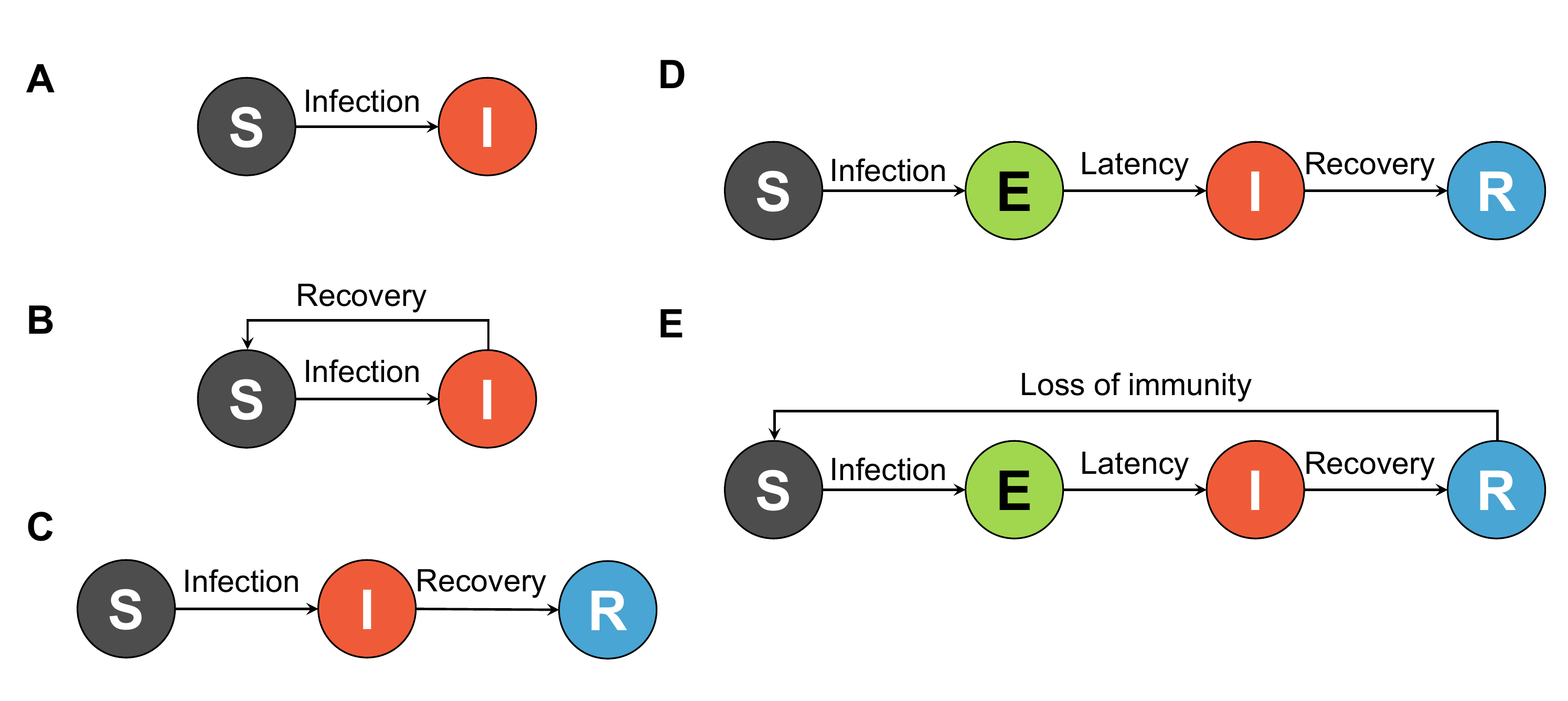}
\caption{Examples of classical compartmental models.}

{\small\noindent
    \parbox{\textwidth}{
(A) The SI model includes only the infection process, where susceptible individuals become infectious. (B) The SIS model extends this by incorporating both infection and recovery, allowing recovered individuals to return to the susceptible state. (C) The SIR model also includes infection and recovery, but recovered individuals remain immune. (D) The SEIR model further introduces a latency period, representing the exposed but non-infectious stage between infection and becoming infectious. (E) The SEIRS model adds loss of immunity, allowing recovered individuals to gradually return to susceptibility, capturing cyclical or endemic disease dynamics.
    }
}

\label{Fig.8}
\end{figure*}

Representative examples include the SI, SIS, SIR, SEIR, and SEIRS models, which have been widely employed to describe the spread of various infectious diseases. In basic SEIR models, the susceptible compartment represents individuals who have not yet been infected but can become infected upon contact with infectious individuals. When a susceptible person is exposed to the pathogen, they move into the exposed compartment, which represents those who are infected but not yet infectious themselves. This stage corresponds to the latent or latent period during which the pathogen replicates within the host but has not yet reached transmissible levels. After the latent period, exposed individuals progress to the infectious compartment, during which they can transmit the disease to others through direct contact or other specified routes, depending on the transmission mechanism. In the end, infectious individuals enter the recovered or removed compartment, where they are assumed to have gained immunity or die. However, if this immunity wanes over time, recovered individuals may lose protection and re-enter the susceptible compartment, thereby completing the epidemiological cycle and enabling the possibility of recurrent outbreaks or the endemic persistence, as shown in \autoref{Fig.8}.

Usually, compartmental models also consider the natural demographic processes, such as births and deaths, to more accurately represent population dynamics over time. For instance, the study \citep{Bjornstad2020} proposed an extended SEIRS model formulated as:
\begin{equation}
\begin{array}{l}
\frac{{{\rm{d}}S}}{{{\rm{d}}t}} = \underbrace {\mu N}_{{\rm{birth}}} - \underbrace {\beta IS/N}_{{\rm{infection}}} + \underbrace {\omega R}_{{\rm{lost immunity}}} - \underbrace {\mu S}_{{\rm{death}}}\\
\frac{{{\rm{d}}E}}{{{\rm{d}}t}} = \underbrace {\beta IS/N}_{{\rm{infection}}} - \underbrace {\sigma E}_{{\rm{latency}}} - \underbrace {\mu E}_{{\rm{death}}}\\
\frac{{{\rm{d}}I}}{{{\rm{d}}t}} = \underbrace {\sigma E}_{{\rm{latency}}} - \underbrace {\gamma I}_{{\rm{recovery}}} - \underbrace {\left( {\mu  + \alpha } \right)I}_{{\rm{death}}}\\
\frac{{{\rm{d}}R}}{{{\rm{d}}t}} = \underbrace {\gamma I}_{{\rm{recovery}}} - \underbrace {\omega R}_{{\rm{lost immunity}}} - \underbrace {\mu R}_{{\rm{death}}}
\end{array}
\text{,}
\label{eq:2}
\end{equation}
where $N$ is the sum population; $\mu$ denotes the natural birth and death rate; $\beta$ is the contact rate between the susceptible and the infectious individuals; $\omega$ represents the rate of immunity loss; $\sigma$ is the progression rate from exposed to infectious; $\gamma$ is the recovery rate; $\alpha$ is the disease-induced death rate. The population remains at $N$ over times, and the associated basic reproduction number is ${R_0} = [\sigma /(\sigma  + \mu )] \times [\beta /(\gamma  + \mu  + \alpha )]$. When $R_0>1$ , both $E$ and $I$ are excepted to approach an endemic equilibrium, with non-zero levels of exposure and infected individuals. When $R_0<1$, the $E$ and $I$ converge to $0$, reaching a disease-free equilibrium.

To effectively parameterize and fit compartmental models, especially when incorporating human mobility patterns, empirical data are indispensable for initial conditions and transition rates definitions \citep{Wang2018}. The key sources include: (1) case reports from public health agencies to estimate the initial susceptible and infected populations as well as recovery or progression rates; (2) vital statistics registries, which provide birth and death rates for compartmental models that account for natural demographic processes; (3) public health interventions (e.g., The Oxford COVID-19 Government Response Tracker \citep{2020}) and (4) human mobility records (see \autoref{sec:2}), including airline and road passenger flow records from various carriers, which are used to model cross-regional transmission dynamics and to capture population variations resulting from non-demographic mobility changes.

A SEIR model integrated with human mobility was proposed in \citep{Gatto2020}. In this model, mobility is represented by the probability of individuals traveling between communities and making contact, thereby simulating how movement across regions influences disease spread. The model divides the population into six distinct compartments: susceptible (S), exposed (E), presymptomatic (P), infected (I), asymptomatic (A), and recovered (R), and it assumes frequency-dependent contact rates, whereby exposure occurs at a rate described by the force of infection. The derivative of susceptible in community $i$ is given as ${\dot S_i} =  - {\lambda _i}{S_i}$ where the corresponding force of infection for the community $i$ is expressed as:
\begin{equation}
{\lambda _i} = \sum\limits_{j = 1}^n {C_{ij}^s} \frac{{\sum\limits_{Y \in \{ P,I,A\} } {\sum\limits_{k = 1}^n {{\beta _Y}} } C_{kj}^Y{Y_k}}}{{\sum\limits_{X \in \{ S,E,P,I,A,R\} } {\sum\limits_{k = 1}^n {C_{kj}^X} } {X_k}}}
\text{,}
\label{eq:3}
\end{equation}
where $C_{ij}^X$ (with $X \in \{ S,E,P,I,A,R\} $ ) is the probability ($\sum _{j = 1}^nC_{ij}^X = 1$ for all $i$ and $X$) that individuals in an epidemiological state $X$ originating from the community $i$ enter into contact with individuals in the community $j$ as either residents or travelers from the community $k$. The term $\beta_X$ represents the specific transmission rates of three infectious classes $X \in \{ P,I,A\} $. 

To date, compartmental models remain among the most commonly used mathematical frameworks in the field of epidemic modeling \citep{Askitas2021,Lv2024}. However, the assumption of homogeneous mixing is often unrealistic in real-world contexts. Empirical studies have shown that human mobility follows a power-law distribution,wherein the majority of individuals engage in short-range movements while only a small proportion travel long distances \citep{Barabasi2005}. This highly heterogeneous mobility pattern translates into non-uniform contact structures, meaning that the probability of interaction between individuals is far from evenly distributed. Furthermore, the studies of scale-free networks \citep{Barabasi2009} and small world networks \citep{Saramaki2005} in human society reinforces this view, demonstrating that epidemic spread is strongly influenced by network topology and mobility heterogeneity.

\subsection{Metapopulation Models}\label{sec:5.2}

Classical compartmental models typically assume a homogeneously mixed population, neglecting spatial heterogeneity and differences in local contact structures, which potentially limits their ability to accurately represent the spatial dynamics of infectious disease spread \citep{Hazarie2021}. To address this limitation, metapopulation models are developed as spatially explicit extensions of compartmental frameworks. These models divide the total population into multiple subpopulations according to cities or communities and explicitly incorporate migration rates to simulate individual movement across them, thereby capturing the heterogeneity in contact opportunities and mobility-driven transmission pathways \citep{Finger2016,Gatto2020,Mou2024a}. 

A classic metapopulation model proposed by Colizza \& Vespignani \citep{Colizza2008} treat a system of $V$ subpopulations connected by a heterogeneous network whose degree distribution is $P(k)$, coupling compartmental infection dynamics (reaction) with population movement across network links. The analysis uses degree-block variables so that quantities averaged over nodes of degree $k$ (e.g. the average subpopulation size $N_k$, susceptible $S_k$, infected $I_k$, and recovered $R_k$) are the basic state variables and network heterogeneity enters through the moments $\langle {k^\alpha }\rangle $. 

At the mobility level the mean-field degree-block equation at time $t$ is written as:
\begin{equation}
{\dot N_k}(t) =  - {p_k}{N_k}(t) + \frac{k}{{\langle k\rangle }}\sum\limits_{{k^\prime }} {{k^\prime }} P({k^\prime }){d_{{k^\prime }k}}{N_{{k^\prime }}}(t)
\text{,}
\label{eq:4}
\end{equation}
where ${d_{{k^\prime }k}}$ is the per-individual diffusion rate from degree-$k$ subpopulations to degree-$k$ subpopulations and ${p_k} = \sum\limits_{{k^\prime }} {{k^\prime }} P({k^\prime }|k){d_{k{k^\prime }}}$ is the total leaving rate from degree-$k$ subpopulations. Two main specifications of the diffusion kernel ${d_{k{k^\prime }}}$ are considered. In the traffic-dependent mobility rate, the flow between nodes is proportional to the product of their degrees, 
\begin{equation}
{d_{k{k^\prime }}} = p{w_0}\frac{{{{(k{k^\prime })}^y}}}{{{T_k}}}
\text{,}
\label{eq:5}
\end{equation}
where $p$ is a global mobility rate, $w_0$ and $y$ are scaling parameters describing how traffic volume scales with degree, and ${T_k} \propto {k^{1 + y}}$ ensures conservation of total flow. This leads to stationary population sizes ${N_k} \propto {k^{1 + y}}$. 
Alternatively, in the population-dependent mobility rate, the diffusion rate depends explicitly on the local population, 
\begin{equation}
{d_{k{k^\prime }}} = {w_0}\frac{{{{(k{k^\prime })}^y}}}{{{N_k}}}
\text{,}
\label{eq:6}
\end{equation}
allowing $N_k$ to be treated as an independent parameter. By combining the infection dynamics with the mobility process, they expressed the full metapopulation model as a set of reaction–diffusion equations for each degree class. For the infectious compartment, this reads:
\begin{equation}
\frac{{d{I_k}}}{{dt}} =  - {p_k}{I_k} + (1 - {p_k})(\beta  - \mu ){I_k} + \frac{k}{{\langle k\rangle }}\sum\limits_{k'} {k'} P(k'),{d_{k'k}}[(1 - \mu ){I_{k'}} + \beta {G_{k'}}]
\text{,}
\label{eq:7}
\end{equation}
where ${G_{k'}} \approx \frac{{{I_{k'}}{S_{k'}}}}{{{N_{k'}}}}$ represents the local reaction term capturing the infection process within subpopulations; $\beta$ represents the infected ratio; $\mu$ represents the recovered ratio. This formulation explicitly couples the internal epidemic dynamics to the external mobility-driven mixing among subpopulations.

Implementing metapopulation models relies on spatially granular data sources to accurately parameterize subpopulation demographics, local transmission dynamics, and inter-region migration flows. Essential inputs include demographic data from census or household surveys (e.g., data from National Statistics Offices), public health surveillance systems (e.g., WHO Global Health Observatory), or openly available data sources (e.g., the WorldPop’s Open Spatial Demographic Data, www.worldpop.org) for initializing compartment sizes and region-specific parameters like $R_0$ or recovery rates. Population migration forms the backbone of interregional transition modeling, sourced from mobile phone records such as CDRs, Google Community Mobility Reports, Meta’s Data for Good portal, or Apple Mobility Trends for near real-time origin–destination fluxes and commuting patterns (see \autoref{sec:2}); transportation logs from airline databases, rail and bus ticketing systems, and ride-sharing GPS traces to quantify long- and short-range migrations; and survey-derived matrices such as the Community Survey to characterize chronic mobility flows. Intra-subpopulation contact behaviors are further informed by social surveys and Bluetooth proximity data from contact-tracing applications.

A representative applied example is the metapopulation-based SIR model developed for the 2010 cholera outbreak in Haiti \citep{Rinaldo2012}. This model explicitly accounted for uneven population mobility between Haitian regions and hydrological mechanisms, such as rainfall-driven runoff, that facilitated the environmental spread of Vibrio cholerae through water systems after the earthquake. Communities were modeled as interconnected nodes, linked by both human migration flows and waterway networks, thereby enabling both direct and indirect transmission across distant regions. This framework highlighted how spatial heterogeneity, particularly power-law mobility patterns, significantly influenced the outbreak’s rapid geographic expansion and provided critical insights for designing more effective public health interventions. 

Mathematically, the model extended the classic SIR equations by introducing inter-patch mobility terms. For a metapopulation consisting of N subpopulations, the dynamics for subpopulation  can be expressed as: 
\begin{equation}
\frac{{d{S_i}}}{{dt}} = \mu ({H_i} - {S_i}) - {{\cal F}_i}(t){S_i} + \rho {R_i}
\text{,}
\label{eq:8}
\end{equation}
\begin{equation}
\frac{{d{l_i}}}{{dt}} = {{\cal F}_i}(t){S_i} - (\gamma  + \mu  + \alpha ){I_i}
\text{,}
\label{eq:9}
\end{equation}
\begin{equation}
\frac{{d{R_i}}}{{dt}} = \gamma {I_i} - (\rho  + \mu ){R_i}
\text{,}
\label{eq:10}
\end{equation}
\begin{equation}
\frac{{d{B_i}}}{{dt}} =  - {\mu _B}{B_i} - I\left( {{B_i} - \sum\limits_{j = 1}^n {{P_{ji}}} \frac{{{H_j}}}{{{H_i}}}{B_j}} \right) + \frac{\rho }{{{H_i}}}\left[ {1 + \Phi {J_i}(t)} \right]\left[ {(1 - m){I_i} + m\sum\limits_{j = 1}^n {{Q_{ji}}} {I_j}} \right]
\text{,}
\label{eq:11}
\end{equation}
where $S_i$, $I_i$, and $R_i$ are the abundance of susceptible, infected, and recovered in node $i$; $H_i$ is the population size of node $i$; $\mu$, $\rho$, $\gamma$, and $\alpha$ are the rates of natural human mortality, immunity wanning, recovery, and disease death, respectively. $B_i$ represents the concentration of Vibrio cholera in the water reservoir at node $i$, the total contact rate ${{\cal F}_i}(t) = \beta \left( {\left( {1 - m} \right){B_i}/\left( {K + {B_i}} \right) + m\sum\limits_{j = 1}^n {{Q_{ij}}} {B_j}/\left( {K + {B_j}} \right)} \right)$ incorporating both local disease transmission within a given node and transmission associated with mobility between nodes. The human mobility patterns are modeled using an Origin-Destination (OD) matrix, wherein individuals depart from their origin node i with probability $m$, arrive at a destination node $j$ with probability $Q_{ij}$, and subsequently return to their home node $i$. To characterize the connectivity between spatial units, a gravity-like model is employed, in which the connection strength $Q_{ij}$ decays exponentially with the distance between nodes:
\begin{equation}
{Q_{ij}} = \frac{{{H_i}{e^{ - \frac{{{d_{ij}}}}{D}}}}}{{\sum\limits_{k \ne i}^N {{H_k}} {e^{ - \frac{{{d_{ik}}}}{D}}}}}
\text{,}
\label{eq:12}
\end{equation}
where $d_{ij}$ represents the shortest path distance between nodes $i$ and $j$, while $D$ denotes the deterrence cutoff distance that controls the rate of decay in interaction strength. 

Both compartmental and metapopulation models are based on the same theoretical framework of compartmental dynamics. In these two types of models, human mobility is primarily used as a basis for parameter estimation, helping to more accurately adjust the population size within each compartment or transmission coefficients between compartments. In general, they do not necessarily capture the full complexity of real-world, topological interaction structures. However, such topological structures, typically represented by networks, serve as critical pathways for epidemic transmission, whether in mobility networks or contact networks. Network-based models enable more flexible representations of transmission processes \citep{Li2025h,Zhang2024d} and allow epidemic simulations to be conducted using network theories such as percolation \citep{Xu2025}, cascade dynamics \citep{Chen2014}, and reaction–diffusion processes \citep{Sun2025c}. These capabilities go beyond the scope of traditional compartmental models.

\subsection{Network-based models}\label{sec:5.3}
Building upon the metapopulation framework introduced earlier, network-informed models offer a more granular and realistic representation of transmission dynamics by explicitly incorporating the heterogeneous structure of human interactions and mobility. Two primary types of data-driven networks are typically employed in the context of epidemics. The first is the mobility network, where nodes correspond to subpopulations or geographic regions, and edges reflect the probabilities or intensities of movement between them. The second is the contact network, which represents individuals as nodes and edges as potential disease-spreading interactions, such as social contact, co-residence, or workplace proximity. These network structures allow for more accurate characterization of transmission pathways, especially in systems where homogeneous mixing assumptions fail to reflect real-world complexity \citep{Moore2020,Wang2024d}.

Another category of epidemic propagation models is network-dynamics models based on algebraic graph theory. These models assume an underlying contact matrix and derive algebraic properties of the network to characterize epidemic network dynamics. This approach circumvents computationally intensive simulations based on real-world data. Instead, by analyzing the system’s dynamic equations, one can infer macroscopic properties of the network and perform direct theoretical analysis of epidemic dynamics \citep{Pare2020}.

\subsubsection{Mobility Network Models}\label{sec:5.3.1}

Mobility networks capture the movement of individuals across different locations, enabling the modeling of how infections propagate through spatially distributed populations. These networks are not static; rather, they evolve over time under the influence of various factors such as social behavior, economic activity, transportation infrastructure, and even public health interventions (e.g., lockdowns, quarantine orders, and travel restrictions) \citep{2021i}. This approach generalizes the traditional compartmental models, wherein individuals are assumed to mix homogeneously within each compartment or location but may migrate between compartments over time, based on empirical or modeled mobility patterns. For example, an epidemic spread across countries via air travel routes can be effectively modeled using this framework, allowing researchers to simulate long-range transmission and assess intervention strategies across borders \citep{Moreno2004}. 

In general, mobility networks have two main applications in epidemic modeling. First, they are often incorporated as part of metapopulation models to support parameter estimation and to simulate spatial coupling between subpopulations. Second, mobility networks can also serve as independent dynamical systems, in which epidemic processes are modeled as cascading or diffusive processes over the network structure.

Balcan et al. proposed a data-driven multiscale framework, combining global air travel and short-range commuting flows within a unified model, Global Epidemic and Mobility (GLEaM) \citep{Balcan2009}. This model constructs its multiscale mobility network entirely from empirical data describing both global air transportation and short-range commuting flows, integrating these within a unified metapopulation framework. GLEaM models two mobility processes operating on different temporal scales. Air travel  occurs on a daily time scale and is implemented explicitly as stochastic movement: each individual in subpopulation $j$ has a probability ${p_{j\ell }} = \frac{{{\omega _{j\ell }}}}{{{N_j}}}$ of traveling to subpopulation $\ell $ on a given day, where ${\omega _{jl}}$ is the traffic flow per unit time, and $N_j$ is the population of node $j$. Commuting mobility, in contrast, operates on a much faster time scale—an average round-trip duration of about one-third of a day. GLEaM uses a time-scale separation approximation, assuming that commuting flows reach a quasi-equilibrium much faster than the disease dynamics evolve. Let ${w_{ji}}$ and ${\sigma _{ji}} = {\omega _{ji}}/{N_j}$ denote the commuting flow and rate from $j$ to its neighbor $i$; ${\sigma _j} = \sum\limits_i {{\sigma _{ji}}} $ is the total outbound commuting rate from $j$, and ${\tau ^{ - 1}}$ is the average return rate. Then, at equilibrium, the number of residents from   present at home and in neighboring subpopulations is given by:
\begin{equation}
{N_{jj}} = \frac{{{N_j}}}{{1 + {\sigma _j}/\tau }}
\text{,}
\label{eq:13}
\end{equation}
\begin{equation}
{N_{ji}} = \frac{{{N_j}({\sigma _{ji}}/\tau )}}{{1 + {\sigma _j}/\tau }}
\text{.}
\label{eq:14}
\end{equation}
These quantities describe the steady-state distribution of individuals between their home location and the neighboring areas they temporarily visit for work or daily activities. The effective force of infection experienced by the susceptible in subpopulation   incorporates both local interactions and those occurring during short-range travel, and is expressed as:
\begin{equation}
{\lambda _j} = \frac{{{\lambda _{jj}}}}{{1 + {\sigma _j}/\tau }} + \sum\limits_i {\frac{{{\lambda _{ji}}({\sigma _{ji}}/\tau )}}{{1 + {\sigma _j}/\tau }}}
\text{,}
\label{eq:15}
\end{equation}
where ${\lambda _{jj}}$ represents the local infection pressure within $j$, and ${\lambda _{ji}}$ represents the infection force acting on individuals from $j$ while visiting neighboring subpopulations $i$. Through this integration of empirical air transportation data, population distribution, and commuting statistics within a mathematically consistent framework, GLEaM establishes a fully data-driven multiscale mobility network.

Assume that each subpopulation remains demographically stable over time and that disease transmission occurs entirely through local interactions within each subpopulation. From the perspective of cascading process over mobility network \citep{Citron2021,Cosner2009,Sattenspiel1995}, population movement between locations can be described as a diffusive process, modeled as follows \citep{Cosner2009}:
\begin{equation}
\frac{{d{N_i}}}{{dt}} =  - \sum\limits_{j = 1}^K {{f_{i,j}}} {N_i} + \sum\limits_{j = 1}^K {{f_{j,i}}} {N_j}
\text{,}
\label{eq:16}
\end{equation}
where ${N_i}$ represents the number of individuals currently located at site $i$, and ${f_{i,j}}$ is the rate at which individuals travel from site $i$ to site $j$, where ${f_{i,i}} = 0$ for all $i$. If visitors are allowed to interact with residents at the destination site and then return to their home location at a fixed rate, the model incorporates a visitor-resident dynamic and can be reformulated as \citep{Cosner2009,Sattenspiel1995}: 
\begin{equation}
\frac{{d{N_{i,i}}}}{{dt}} =  - \sum\limits_{j = 1}^K {{\phi _{i,j}}} {N_{i,i}} + \sum\limits_{j = 1}^K {{\tau _{i,j}}} {N_{i,j}}
\text{,}
\label{eq:17}
\end{equation}
\begin{equation}
\frac{{d{N_{i,j}}}}{{dt}} =  - {\tau _{i,j}}{N_{i,j}} + {\phi _{i,j}}{N_{i,i}}
\text{,}
\label{eq:18}
\end{equation}
where an individual from $i$ who is currently located at $j$ will be counted as belonging to the $N_{i,j}$ population, and the number of individuals whose home is $i$ remains constant over time, even if members visit other locations. The constant ${\phi _{i,j}}$ represents the rate at which individuals whose home is $i$ travel to $j$, while the constant ${\tau _{i,j}}$ is the rate at which individuals visiting $j$ from $i$ return home to $i$. 

In \citep{Schlosser2020}, the weekly mobility networks utilizes empirical trip data to construct time-resolved mobility networks ${G_T}$ for each calendar week $T$, capturing dynamic changes in population flows. The edge weights ${w_{ji}}(T)$ are then calculated as the average daily number of trips between counties during this week:
\begin{equation}
{w_{ji}}(T) = {\left| {{D_T}} \right|^{ - 1}}\sum\limits_{{t^\prime } \in {{\cal D}_T}} {{F_{ji}}} ({t^\prime })
\text{,}
\label{eq:19}
\end{equation}
where ${D_T}$ denotes the set of days in calendar week $T$, and ${F_{ji}}(t)$ denotes the migration people from $j$ to $i$ in time $t$. 

To investigate how the global reduction of mobility affects the observations in comparison to structural changes, it constructs rescaled networks $G_{10}^*(T)$ by scaling the weights of the pre-lockdown network of calendar week ten by the flow lost during the week $T$, i.e., 
\begin{equation}
w_{ji}^ \star (T) = {w_{ji}}(T) \times \frac{{\sum\limits_{i,j = 1}^m {{w_{ji}}} (T)}}{{\sum\limits_{i,j = 1}^m {{w_{ji}}} (T = 10)}}
\text{.}
\label{eq:20}
\end{equation}

By constructing mobility network, researchers can incorporate human mobility into epidemic modeling and analyze various lockdown-induced changes in mobility. For instance, during the initial phase of the pandemic, studies found a considerable reduction of mobility in Germany, similar to what was previously reported for other countries that passed and implemented comparable policies \citep{Cosner2009,Sattenspiel1995,Schlosser2020}.

\subsubsection{Contact Network Models}\label{sec:5.3.2}
A growing body of research has focused on characterizing contact patterns by developing data-informed models that explicitly integrate human mobility within epidemic transmission dynamics. The data sources used to construct contact networks usually include cellular signaling, satellite positioning, IP and Wi-Fi location tracking, and IoT location tracking (see \autoref{sec:2}) \citep{Bansal2010}. Contact network models emphasize the heterogeneity of infection risk arising from both individual-level interactions and mobility-induced exposure, thereby enhancing the accuracy and predictive power of epidemic forecasts. Generally, contact network models can be regarded as a subtype of agent-based models (ABMs). ABMs simulate individuals or entities (e.g., persons, communities, or cities) as agents with personal characteristics and decisions, while contact network models focus specifically on the topology of interpersonal interactions that drive transmission. This topological emphasis allows researchers to examine how heterogeneous contact structures (e.g., age structure) affect epidemic spread, a perspective that is often less explicit in traditional ABMs.

Here we present two representative modeling approach based on data-informed contact networks: the temporal-evolving contact network model, which describes the dynamics formation and dissolution of contacts over time \citep{Bansal2010}, and the scenario-based multilayer contact network model, which characterizes different contact types across multiple social or behavioral layers \citep{Liu2018}.

Machens et al. \citep{Machens2013} constructed several empirical contact network models from high-resolution human proximity data. In these models, infection transmission occurs whenever a susceptible and infectious individual are in contact, with an instantaneous transmission rate $\beta$, giving an infection probability of $1 - {e^{ - \beta T}}$ over a contact duration $T$. A time-aggregated version, the heterogeneous static network, replaces the temporal structure with weighted static edges, where the weight ${w_{ij}} = \sum\limits_t \Delta  {t_{ij}}(t)$ represents the total contact time between individual $i$ and $j$. To move from individuals to role-based mixing patterns, they defined class-level contact matrices. For class $X$   and $Y$ with $N_X$ and $N_Y$ individuals and total cumulative contact time ${W_{XY}}$, each pair $(x,y)$ with $x \in X$, $y \in Y$ is assigned a uniform weight:
\begin{equation}
{w_{XY}} = \frac{{{W_{XY}}}}{{{N_X}{N_Y}}}(X \ne Y)
\text{,}
\label{eq:21}
\end{equation}
and ${w_{XX}} = \frac{{{W_{XX}}}}{{{N_X}({N_X} - 1)/2}}$ for within-class pairs.

Complementing this temporal perspective, the scenario contact network model developed in \citep{Liu2018} provides a flexible framework to assess the measurability and relevance of classical epidemic indicators, such as the basic reproduction number  and generation time \citep{Biggerstaff2014}. This model is structured as a multi-layer contact network, with each layer representing a distinct type of social interaction. Additionally, demographic stratification including attributes such as sex, age, and occupation can be incorporated as layered subpopulations \citep{Salje2021}. This stratification introduces both visible and invisible barriers to transmission, thereby enabling the modeling of targeted intervention strategies for specific demographic groups. The core methodology in \citep{Liu2018} involves the construction of multiplex contact networks, which represent interactions within households, schools, workplaces, and the broader community. Each layer captures a specific interaction context, allowing for a more realistic representation of contact patterns. The node degree distribution aligns with empirical data, and the weighted number of contacts for each individual can be computed as follows:
\begin{equation}
\sum\limits_{l \in \left\{ {h,s,w,c} \right\}} {{w_l}} {c_l}(i)
\text{,}
\label{eq:22}
\end{equation}
where ${c_l}(i)$ is the number of edges that node $i$ has within layer $l$, and $w_i$ is the weight of layer $l$ with $\sum\limits_{h,s,w,c} {{w_l}}  = 1$.

Besides the data-informed contact network models reviewed above, another line of research investigates epidemic spreading on contact networks through the lens of algebraic graph theory. This approach focuses on the algebraic properties of the adjacency matrix $A$, where ${A_{ij}} > 0$ indicates that node $i$ and $j$ are connected, and the value of ${A_{ij}}$ represents the strength of their connection (e.g., contact frequency or strength). Algebraic graph theory explores the relationships between the algebraic characteristics of $A$, the structural features of the corresponding graph, and the resulting network behavior, providing a framework for describing human mobility via adjacency matrices.

This class of models was firstly introduced by Lajmanovich and Yorke \citep{Lajmanovich1976}, who proposed the SIS model on networks. Later, Mieghem et al. \citep{VanMieghem2009} proved that this deterministic network model corresponds to a mean-field approximation of an exponential-dimension Markov chain representation of the stochastic SIS process. The general network SIS model is given as:
\begin{equation}
{\dot x_i}(t) = \beta (1 - {x_i}(t))\sum\limits_{j = 1}^n {{a_{ij}}} {x_j}(t) - \gamma {x_i}(t)\quad i = (1, \ldots ,n)
\text{,}
\label{eq:23}
\end{equation}
where ${x_i}(t)$ is the infection proportion of node $i$, $a_{ij}$ is the connection strength between node $i$ and node $j$, $\beta$ and $\gamma$ are infection rate and recovery rate, respectively, which are assumed to be uniform across all nodes. The system can be expressed in vector form as:
\begin{equation}
\dot x(t) = \beta ({E_n} - {\rm{diag}}(x(t)))Ax(t) - \gamma x(t)
\text{,}
\label{eq:24}
\end{equation}
where $x(t) = ({x_1}(t), \ldots ,{x_n}(t))$ is the infection vector at time $t$, $E_n$ is the $n$-dimensional identity matrix, and ${\rm{diag}}(x(t))$ is a diagonal matrix whose diagonal is $x(t)$, and $A$ is an irreducible adjacency matrix whose elements are $a_{ij}$, and the corresponding graph is strongly connected. For such models defined by the adjacency matrix, there exists an epidemic threshold given by:
\begin{equation}
\beta {\lambda _{\max }}(A)/\gamma
\text{,}
\label{eq:25}
\end{equation}
where ${\lambda _{\max }}(A)$ denotes the dominant eigenvalue of $A$ which equals to the spectral radius $\rho (A)$. When the threshold $\beta {\lambda _{\max }}(A)/\gamma $ is less than $1$, the disease will eventually die out and the system will converge to the Disease-Free Equilibrium (DFE), whose region of attraction covers all possible initial conditions. When the threshold $\beta {\lambda _{\max }}(A)/\gamma $ is greater than $1$, two equilibria exist: DFE and a unique non-zero endemic equilibrium. The DFE is unstable, while the endemic equilibrium is locally exponentially stable, with its region of attraction being all possible initial conditions except DFE. This threshold is referred to as the basic reproduction number $R_0$, analogous to the reproduction number in other epidemic models. The detailed proofs of these results can be found in research by Lajmanovich and Yorke \citep{Lajmanovich1976}, Fall et al. \citep{Fall}, Khanafer et al. \citep{Khanafer2016}, and Mei et al. \citep{Mei2017}.

Similarly, by performing Euler’s method to the continuous-time system or by applying probabilistic modeling directly to real epidemic processes followed by specialization and truncation, a discrete-time network SIS model can be derived \citep{Pare2020a}. Contact Network SIR model is another widely used class of epidemic models. A general network-based SIR model is given by:
\begin{equation}
\begin{array}{l}
{{\dot x}_i} =  - {x_i}\sum\limits_{j = 1}^n {{A_{ij}}} {y_j}\\
{{\dot y}_i} = {x_i}\sum\limits_{j = 1}^n {{A_{ij}}} {y_j} - \gamma {y_i}
\end{array}
\text{,}
\label{eq:26}
\end{equation}
where $i = 1, \ldots ,n$; $x_i$ is the susceptible proportion of node $i$; $y_i$ is the infection proportion of node $i$; $\gamma$ is the removed/recovery rate, and the infection rate is normalized to $1$. It can be described in following vector form:
\begin{equation}
\begin{array}{l}
\dot x =  - {\rm{diag}}(x)Ay\\
\dot y = {\rm{diag}}(x)Ay - \gamma y
\end{array}
\text{,}
\label{eq:27}
\end{equation}
where $x = {({x_1}, \ldots ,{x_n})^T}$, $y = {({y_1}, \ldots ,{y_n})^T}$, and $A$ is the adjacency matrix. In \citep{Alutto2025}, Alutto et al. focused on analyzing the dynamical behavior of a rank-1 adjacency matrix, which corresponds to the case where all infected individuals are fully mixed across nodes and infect at heterogeneous rates. They derived the invariant set of the dynamics, the explicit expressions of the equilibria, and the necessary and sufficient conditions for equilibrium stability. Guo et al. \citep{Guo2008} further investigated a network-based SIR model with birth and death rates, derived the threshold of endemic, and, through a detailed Lyapunov analysis, proved the dynamical properties of the system above the threshold.

In addition to studying the dynamics of single-virus infections, Liu et al. \citep{Liu2019a} investigated a distributed continuous-time bi-virus model, in which two viruses compete for infection over the same algebraic network and different networks, leading to distinct convergence behaviors, and the convergence and equilibria analysis based on this model was also analyzed in \citep{Ye2022}. Other related methods for dynamical analysis, prediction, and control are reviewed in the comprehensive surveys by Nowzari et al. \citep{Nowzari2016} and Zino \& Cao \citep{Zino2021}.

Overall, mobility and contact network models, as well as network models based on algebraic graph theory are increasingly utilized in epidemiological research to elucidate the complex, nonlinear dynamics of disease transmission. By explicitly representing the structure and temporal evolution of interactions among nodes, these models offer high-resolution depictions of epidemic spread \citep{Pastor-Satorras2015}. However, these network-based approaches also face several notable challenges. Their accuracy depends critically on the availability and quality of data describing contact and mobility behaviors—data that are often scarce, noisy, or difficult to collect at scale. In addition, simulating large-scale, multilayered networks can be computationally demanding, requiring significant computing resources and optimization techniques. Finally, model validation remains a persistent difficulty due to the dynamic and context-dependent nature of real-world contact structures, which may evolve over time or differ across populations and settings.

\subsection{Agent-Based Models}\label{sec:5.4}
In the context of epidemic modelling, Agent-Based Models (ABMs) typically represent individuals, communities, or even entire cities and countries as autonomous agents with defined specific attributes and decision-making rules. By explicitly encoding heterogeneity in demographic characteristics, behavioral tendencies, and movement or contact patterns, ABMs can achieve a relatively precise simulation of the disease transmission process. ABMs are not limited to predefined network structures and allow for more flexible representations of transmission dynamics. In the following, we introduce two main types of ABM modeling approaches: rule-based models and stochastic models.
\subsubsection{Rule-Based Agent Models}\label{sec:5.4.1}

Epidemic models such as compartmental models and network-based models, often simplify human behavior by assuming homogeneous mixing within groups or by downplaying the heterogeneity of individual actions. These approaches tend to emphasize macro-level state transitions, such as shifts between susceptible and infectious populations, while neglecting the nuanced interactions that occur at the individual level. However, human mobility behavior is inherently stochastic, diverse, and often guided by personal preferences or aversions. Failing to account for individual autonomy in movement and decision-making may severely limit a model’s ability to simulate disease transmission dynamics accurately \citep{Eubank2004}. Integrating human mobility with ABMs can help address the limitations of traditional approaches in capturing individual behavioral dynamics \citep{Faucher2022,Mansilla2001,Pei2021}. ABMs simulate the behaviors and interactions of individual agents, each representing a unique member of the population with independently evolving states over time. These highly flexible models are capable of incorporating detailed mobility patterns, enabling fine-grained, micro-level simulations of disease transmission processes. Human mobility data sources for ABMs typically consist of individual-level mobility data, such as satellite positioning and IoT-based location tracking (see \autoref{sec:2.3} to \autoref{sec:2.6}).

Agent-based models (ABMs) have been widely used to assess how mobility restrictions and behavioral changes influence infection rates (see \autoref{sec:7}). Traditional ABMs operate based on predefined rules, simulating movement behaviors and interactions at a high level of detail \citep{Han2014,Kerr2021,Longini2005,Min2011}. In particular, reference \citep{Cuevas2020} proposed a rule-based ABM designed to evaluate the disease transmission risks in facilities, aiming to capture the spatiotemporal dynamics of epidemic spread. In this model, agents are categorized into two distinct types that make decisions according to specific rule sets—Type A and Type B, corresponding to Rule I and Rule II, respectively. Each agent is assigned a unique mobility pattern, enabling precise simulation of daily activities and interpersonal interactions. These interactions are characterized by various health states, such as susceptible, exposed, infectious, and recovered, which evolve based on predefined rules and interactions with other agents. The rules are aligned with the spatial movement patterns and infection conditions of the agents to represent the transmission process accurately. Furthermore, each agent is assigned a personal profile that includes key social attributes and health status, which govern behavioral responses during interactions. These individual-level features play a critical role in shaping agents' interaction behaviors \citep{Cuevas2020,Zhao2023a,Zhu2022a}.

A recent example of applying agent-based modeling (ABM) to epidemic simulation involves representing individuals as agents who move within urban areas, travel between cities, and interact in various public settings \citep{Atti2008}. This model tracks how mobility restrictions and behavioral adaptations (e.g., social distancing) impact infection rates over time. Agents’ mobility patterns determine the likelihood of contact between individuals, thereby directly influencing the dynamics of disease transmission. Another study investigating how objective mobility affects epidemic spread found that increased mobility significantly raises the final number of infections. For mobile individuals, infection rates at nodes were found to be proportional to their betweenness centrality, while for non-mobile individuals, infection rates were approximately proportional to nodal degree \citep{Tang2009}. These ABM frameworks focus on the core mechanisms of epidemic spread, particularly human behaviors such as movement, dwelling, and social interactions, enabling them to capture the complexity of human mobility and interaction more effectively than aggregate models. Each agent is assigned a mobility trajectory that dictates their daily routines, interactions, and transitions between locations. Decisions such as whether to visit a public venue or remain at home directly affect their exposure risk. Broader patterns of human movement influence the probability of agent contact, thus shaping the overall dynamics of transmission.

\subsubsection{Stochastic Agent Models}\label{sec:5.4.2}

In contrast to rule-based agent models, where agent behaviors follow explicitly defined interaction protocols, stochastic agent-based models introduce inherent randomness and unpredictability in agent interactions, better capturing the complexity and variability observed in real-world social dynamics \citep{2025b,2024p,Gallotti2016}. 

In \citep{Hoertel2020}, it develops a stochastic agent-based microsimulation model is developed to evaluate the potential effects of non-pharmaceutical interventions (NPIs) on the trajectory of the COVID-19 epidemic in France. Each agent in the model represents an individual characterized by demographic and health attributes drawn from national statistics, including age, sex, household composition, and comorbidities known to increase the risk of severe SARS-CoV-2 infection. The synthetic population was constructed to reflect the real-world distribution of risk factors in France, with approximately 36.4\% of the population categorized as vulnerable due to age (over 65 years) or pre-existing medical conditions. The agents interact within a dynamic social contact network, which simulates daily activities such as work, school, family gatherings, public transport, and grocery shopping. Each interaction is parameterized by distance, duration, and frequency, thereby generating probabilistic exposure events. The disease model is overlaid on this network, allowing for infection to spread across contact edges according to a transmission probability function that decays with physical distance and is modulated by preventive behaviors such as mask-wearing and physical distancing. By integrating epidemiological parameters, the model simulates disease progression on an individual level, accounting for the heterogeneity of both exposure risk and clinical severity. Crucially, the stochastic nature of the model enables it to capture the probabilistic and emergent behavior of an epidemic in a heterogeneous population.

\begin{figure*}[htbp]
\centering
\includegraphics[width=0.8\textwidth]{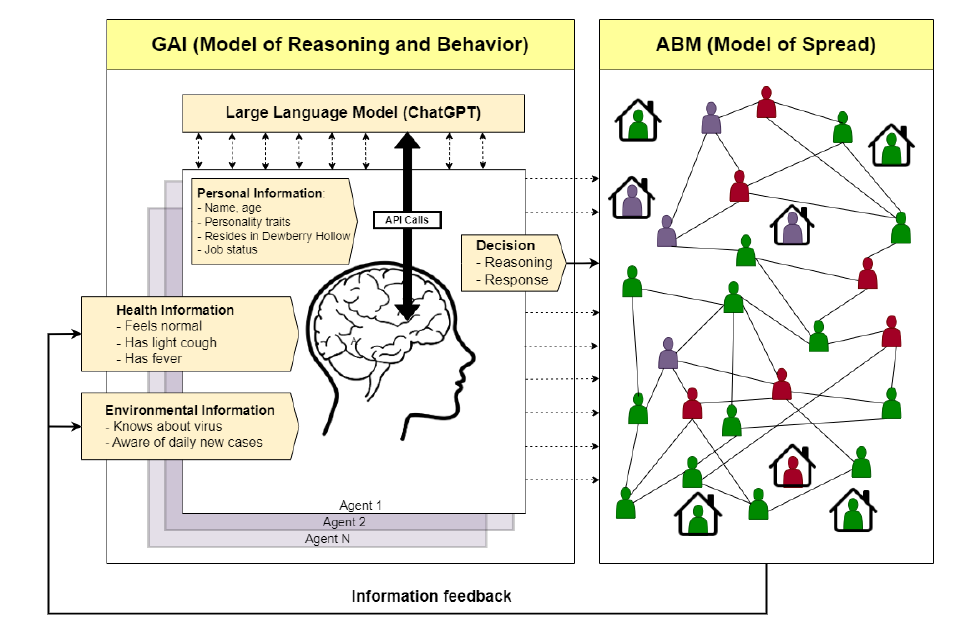}
\caption{Generative agent-based modeling for spreading behaviors.}

{\small\noindent
    \parbox{\textwidth}{
Generative agent-based modeling provides reasoning and decision-making ability for each agent through closed feedback between an AI-based large language model of reasoning (GAI) and a simulation model of spread (ABM). This figure is a reproduction of Fig. 1 in Ref. \citep{Williams2023}.
    }
}

\label{Fig.9}
\end{figure*}

Recent advancements further enhance these models by integrating real-world modeling technologies, such as generative agents with Large Language Models (LLMs). \autoref{Fig.9} demonstrates a framework to model spreading behaviors using generative artificial intelligence \citep{Williams2023}. By employing LLMs, researchers can more realistically simulate individual decision-making processes, social behaviors, and communication patterns, substantially improving the fidelity of epidemic simulations. Such integration allows stochastic agent models to more effectively reflect the unpredictability inherent in human behavior, thus offering more accurate and actionable insights into disease spread dynamics and potential interventions \citep{Tang2009}.

The strength of ABMs lies in their ability to simulate individual behaviors and interactions, providing detailed insights into disease transmission that are often unattainable through more aggregated modeling approaches. However, the inherent diversity and unpredictability of human behavior pose significant challenges to accurately simulating real-world scenarios. Detailed ABMs can be computationally intensive, but advances in computational technology have increasingly enabled large-scale and efficient simulations. By integrating human mobility with ABM techniques, researchers can develop powerful epidemic modeling tools that effectively capture the complexity of human behavior and movement.

\subsection{Machine Learning Models}\label{sec:5.5}
\begin{figure*}[htbp]
\centering
\includegraphics[width=\textwidth]{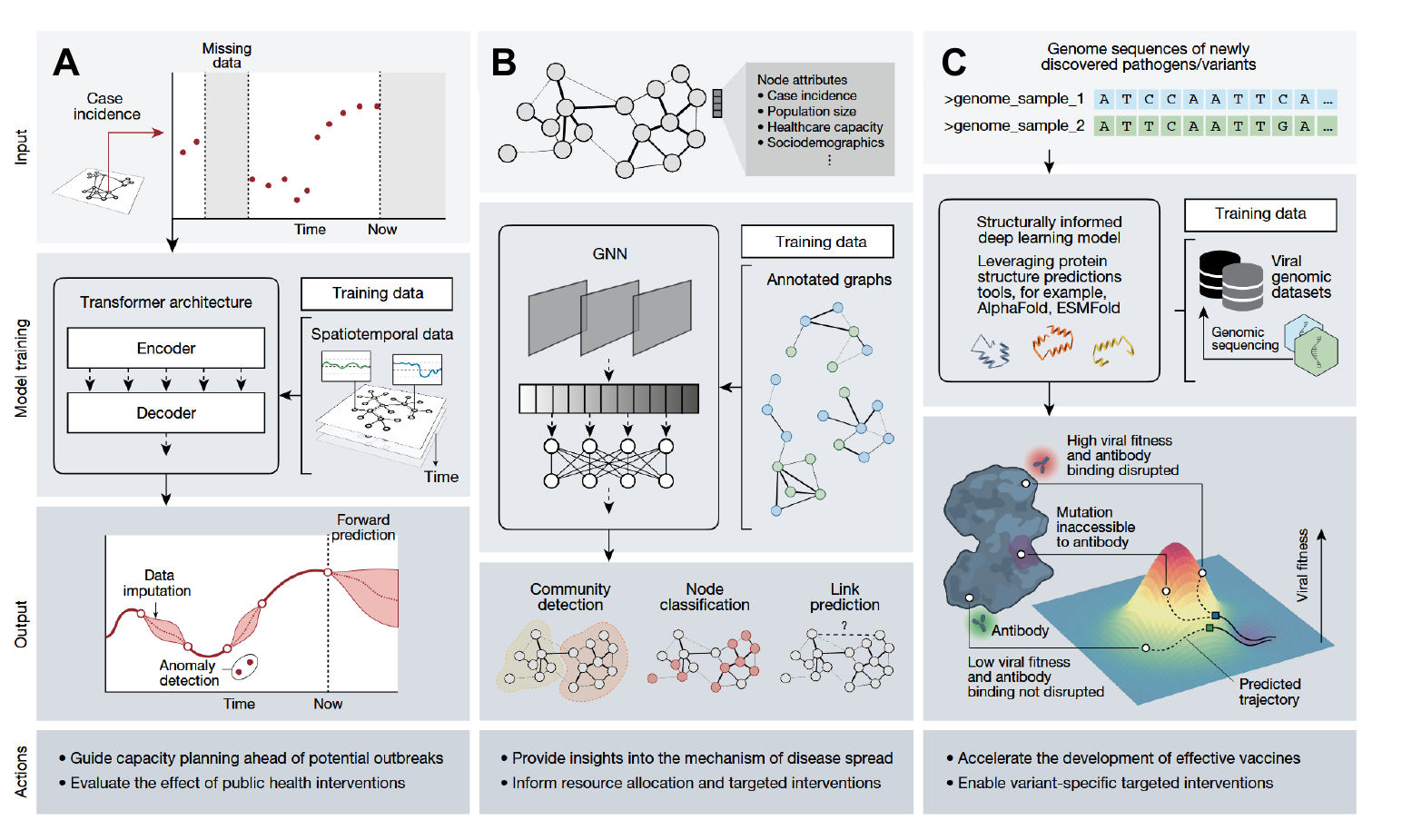}
\caption{Machine learning approaches to tackling key epidemiological questions.}

{\small\noindent
    \parbox{\textwidth}{
(A) By pre-training large-scale foundation models with transformer architecture using time-series epidemic data—whether empirical or simulated—it becomes possible to apply these models in a zero-shot manner for tasks such as forward prediction, data imputation, and anomaly detection. The outputs generated by these models can be utilized to assess the impact of public health interventions and inform capacity planning for future outbreaks. (B) Modeling the spread of infectious diseases with Graph Neural Networks (GNNs) involves representing pathogen transmission through annotated graphs. In these graphs, nodes represent locations or individuals, and edges symbolize potential transmission routes, such as human or vector interactions. Each node is linked to a set of features, like case incidence and population size, which serve as indicators or drivers of disease spread. GNNs can learn intricate patterns from this data, facilitating tasks such as node classification (predicting disease prevalence), community detection (identifying infection clusters), and link prediction (uncovering hidden transmission pathways). These insights offer a deeper understanding of the mechanisms behind disease spread and support efficient resource allocation and targeted interventions. (C) Leveraging biologically informed deep learning models to predict immune-escape mutations involves utilizing recent advancements in protein structure prediction, like AlphaFold and ESMFold. These models can help identify pathogens or variants that are prone to developing mutations that disrupt antibody binding, potentially leading to resistance against current vaccines or therapeutics. Such predictive capabilities can assist in the development of next-generation vaccines and help prioritize containment strategies aimed at emerging variants. This figure is a reproduction of Fig. 1 in Ref. \citep{Kraemer2025}.
    }
}

\label{Fig.10}
\end{figure*}

With advancements in computational and statistical technologies, machine learning (ML) models have been extensively explored and applied in infectious disease modeling \citep{Rodriguez2024,Venkatramanan2021,Ward2022}. ML techniques offer new avenues for predicting contagious disease spread and evaluating intervention strategies \citep{Kraemer2025,Wang2024b}, as shown in \autoref{Fig.10}. These models can handle large, complex datasets and enhance the real-time response capabilities. Traditional ML models, such as Bayesian \citep{Carella2022}, Autoregressive Integrated Moving Average (ARIMA) \citep{Silva2021}, Support Vector Machine (SVM) \citep{Sardar2022}, and Ensemble Learning (EL) \citep{Barlacchi2017}, provide clear insights into how human mobility influences disease transmission, while balancing predictive performance with interpretability. In addition, deep learning models such as Deep Neural Networks (DNNs) \citep{Liu2022}, Graph Neural Networks (GNNs) \citep{Liu2023c}, and Long Short-Term Memory (LSTM) \citep{2021}, have also demonstrated strong performance in epidemic prediction tasks due to their ability to capture complex nonlinear patterns. However, their lack of interpretability and reliance on large-scale data can limit their applicability in some policy-sensitive contexts. Recently, large language models (LLMs) have emerged as powerful tools for extracting knowledge from unstructured data, such as scientific literature, policy documents, and social media. While not directly used for mechanistic modeling, LLMs can support epidemic analysis by synthesizing insights, generating hypotheses, and enhancing decision-making workflows \citep{Du2025}.

\subsubsection{Traditional Machine Learning Models}\label{sec:5.5.1}

Traditional machine learning models for epidemic prediction often rely on aggregated historical mobility statistics (e.g., air, rail, and road travel flows) to model how human movement \citep{2021h} drives cross-regional epidemic transmission dynamics (see \autoref{sec:2.2} and \autoref{sec:2.3}). For instance, Bayesian models are a prevalent forecasting tool, enhancing accuracy by combining mobility data with prior insights into mobility or epidemiology. Within this framework, the prior distribution represents initial parameter assumptions, updated by observed human mobility data into the posterior distribution for final epidemic parameter estimation. By treating mobility probabilistically, these models gauge the likelihood of cross-regional transmission, leading to more accurate identification of infection pathways and emerging hotspots. For instance, a conditional Batesian spatial modeling framework was offered to assess perceived infection risk, modeling the relationship between county-level demographic, socioeconomic, and business attributes across the U.S. \citep{Carella2022}. This approach treats mobility patterns as probabilistic inputs to model transmission likelihood across regions. Conditional on the selected covariates, the percentage in mobility is assumed to follow a Beta distribution:
\begin{equation}
\mathbb{E}[y\mid\bf{x}]\sim B(\mu(\bf{x}),\phi)
\text{,}
\label{eq:28}
\end{equation}
where $\mathbf{x}$ is the set of fixed covariates, given by the selected principal component constructed in Principal Component Analysis and the epidemiological covariate; $\phi $ is the precision parameter of the Beta distribution and $\mu ({\bf{x}})$ its mean, linked to the linear predictor $\mu ({\bf{x}})$ by the default logit-link. Such integration enables the Bayesian model to trace how covariates modulate mobility behaviors and propagate transmission risk. Beyond improving hotspot and pathway detection accuracy, the probabilistic treatment explicitly quantifies model uncertainty, transforming it into a decision-support tool for public health planning. 

As time-series techniques, ARIMA models identify trends, cycles, and seasonality in human mobility. When extended to ARIMAX with mobility as an exogenous regressor, they assess both the impact of population movement on disease spread and temporal dependencies in transmission. The models' interpretable parameters clarify mobility-infection linkages. This study further develops the framework by: (1) analyzing mobility-infection correlations, and (2) introducing a hybrid EEMD-ARIMAX forecasting method where mobility serves as the exogenous input \citep{Silva2021}. The parameters of the $ARIMAX(p, d, q, n)$ model are: $p$, the number of autoregressive terms; $d$, the number of nonseasonal differences needed for stationary; $q$, the number of lagged forecast errors in the prediction equation; $n$, the number of exogenous variables. Besides, $\eta $ (a constant), ${\phi _i}$ ($i = 1, \ldots ,p$), ${\theta _j}$ ($j = 1, \ldots ,q$), and ${\zeta _l}$ ($l = 1, \ldots ,n$) are the parameters of the model. Mathematically, this model can be formulated as:
\begin{equation}
{W_t} = \eta  + \sum\limits_{i = 1}^p {{\phi _i}} {W_{t - i}} - \sum\limits_{j = 1}^q {{\theta _j}} {e_{t - j}} + \sum\limits_{l = 1}^n {{\zeta _l}} {Y_l}
\text{,}
\label{eq:29}
\end{equation}
where ${W_t}$ and ${W_{t - i}}$, for $i = 1, \ldots ,p$, are the predicted values of the time series; ${Y_l}$, for $l = 1, \ldots ,n$, are the exogenous variables of human mobility; and ${e_{t - j}}$, for $j = 1, \ldots ,q$, represent the error terms.  This model incorporates external inputs into the ARIMA model to forecast the number of confirmed cases in epidemics. The external variables encompass meteorological data and human mobility data. By integrating these variables into the ARIMAX model, we can more accurately predict the epidemic development trend of different cities and enhance the prediction accuracy by combining the empirical mode decomposition (EMD) method.

In the human mobility and epidemic modeling context, Support Vector Machines (SVMs) are supervised learning algorithms used for classification and regression tasks. In infectious disease modeling, SVM can classify regions or periods based on the risk level of disease spread by utilizing features extracted from human mobility data, such as travel frequency, distance, and connectivity. On the other hand, Ensemble learning methods (e.g., Random Forests) combine multiple models to enhance predictive performance. They are especially adept at handling large datasets with numerous features, making them suitable for analyzing complex human mobility patterns. Random Forests can identify critical predictors of disease spread by evaluating the contribution of different mobility factors. These models provide feature importance scores, which are interpretable and help clarify which mobility factors most significantly influence transmission dynamics. Here is research where SVM and Random Forests are used to identify everyday activities between infected and uninfected individuals to help stop the spread of pandemic diseases \citep{Sardar2022}. Another research focuses on classifying whether an individual will exhibit flu-like symptoms based on their mobility pattern. The Random Forest classifier predicts symptom presence, focusing on short-term mobility behaviors by analyzing features such as the number of different places visited, the total displacement, and other movement-related data. The model achieves notable performance, with an Area Under the Curve (AUC) score of 0.57 and an F1-score of 0.77, demonstrating its ability to identify key mobility features that signal impending symptoms \citep{Barlacchi2017}.

Both SVM and Random Forests can effectively incorporate human mobility data, such as travel frequency, distance, and connectivity, to classify regions or periods based on the risk of disease spread. These models can also identify critical predictors of transmission by evaluating various mobility factors, such as travel hubs, transportation modes, and population density. It helps assess and mitigate transmission risks. For example, datasets like Google Community Mobility Reports, SVM, and Random Forests can correlate movement patterns with infection rates, providing valuable insights for controlling disease outbreaks and identifying which aspects of mobility have the most significant impact on transmission dynamics.

\subsubsection{Deep Learning Models}\label{sec:5.5.2}

Deep learning models are growing prominent in mobility-aware epidemic forecasting \citep{Yang2025c,Wang2024b}. Their foundation lies in multilayered neuron assemblies: individual neurons apply trainable weight parameters to incoming signals before outputting transformed results via nonlinear activations.

High-resolution spatiotemporal trajectory data, such as GPS traces, mobile phone signaling, and IoT location records (see \autoref{sec:2.3} to \autoref{sec:2.6}), allow models like LSTM or DNN to learn sequential patterns in individual or collective movement. Moreover, network-structured mobility data, represented as graphs where nodes denote geographic units and edges represent population flows, are utilized in Graph Neural Networks (GNNs) to model spatial interactions and contagion pathways. These data sources enable deep learning models to achieve high predictive accuracy and spatial-temporal granularity in epidemic forecasting. For instance, DNNs can predict infection rates by processing high-dimensional features derived from human mobility data, such as GPS traces. These networks mimic the human brain's neural connections, allowing them to model intricate, non-linear relationships in large datasets. They can identify hidden patterns that are difficult for traditional models to detect. Study \citep{Liu2022} proposed an IPSO-DNN model to predict social distancing efficacy through mobility pattern analysis. This model consists of three key stages, namely, data preprocessing, IPSO-DNN hyperparameter optimization, and model evaluation. Human mobility data is scaled and split into training and testing sets, ensuring features are normalized to improve model learning. Then, the IPSO algorithm minimizes the mean squared error of the DNN’s prediction on training data. At last, the optimized DNN model predicts the impact of social distancing based on the testing data. The model quantifies how social distancing reduces transmission rates and confirms its critical role in pandemic containment. Furthermore, intervention duration and compliance intensity significantly influence epidemic trajectories.

GNNs are particularly well-suited for epidemic modeling due to their ability to work with graph-structured data such as the OD matrix and contact network. The features of human mobility networks make GNNs ideal for modeling the interconnectedness of such systems. However, in real-world situations, strict privacy data protection regulations result in severe data sparsity problems (i.e., limited case and location information). To address these challenges of data sparsity, researchers propose a Deep Graph Diffusion Infomax (DGDI) from the micro perspective mobility modeling to compute the relevance score between a diffusion and a location \citep{Liu2023c}. From an alternative standpoint, researchers have put forth a network-based deep learning approach to address the changes in spatiotemporal travel mobility and community structure detection induced by the pandemic, which is based on the premise of normalized preparedness. By jointly optimizing graph learning and network analysis in an end-to-end system, this approach models evolving transportation networks. The resulting complex network metrics reveal spatiotemporal mobility transformations and statistical dependencies between travel modes. The findings reveal a reduction in connectivity and travel diversity across various modes, with post-pandemic recovery characterized by polycentric structures and increased bike-sharing usage \citep{Chang2024}. These GNN models spatial and structural interconnectivity within mobility networks, identifying complex contagion dynamics across population clusters through learned relational dependencies. Their ability to model changes in travel behavior and community structures during a pandemic makes them particularly useful for developing targeted intervention strategies and understanding the long-term impacts of mobility shifts on epidemic dynamics.

LSTM networks are effective for time-series forecasting in epidemic modeling, as they can process sequences of mobility data and infection rates, capturing patterns and trends in human movement and infection spread. By learning these dynamics over time, LSTMs enable more accurate predictions of future outbreaks and support timely interventions. This feature of LSTM is helpful for rapidly probing and quantifying the effects of government interventions, such as lockdown and reopening strategies. This study presents a deep learning framework based on LSTM networks for epidemiology system identification from noisy and sparse observations with quantified uncertainty, which is trained on Google and Unicast mobility data \citep{2021}. The method integrates graph learning and optimization to model travel networks dynamically, uncover changes in user mobility patterns, and explore the relationships between different travel modes. The findings reveal significant trip volume and connectivity reductions during the pandemic, with post-pandemic recovery showing a shift towards more polycentric travel patterns and increased use of bike-sharing services. By capturing temporal patterns and trends in human movement, LSTMs model the non-linear relationships between past mobility behaviors and current infection levels, providing accurate predictions even when dealing with complex, time-dependent data. These networks are instrumental in assessing how changes in human mobility such as those caused by lockdowns or public health interventions affect the trajectory of an epidemic. However, while LSTMs offer predictive solid performance, their internal mechanisms, like gates controlling information flow, are complex and hinder interpretability. This trade-off between model transparency and predictive accuracy remains challenging, as LSTMs focus more on accurate forecasting than explanation.

\subsubsection{Large Language Models}\label{sec:5.5.3}
In recent years, large language models (LLMs) have catalyzed a pivotal transformation in the field of natural language processing. LLMs can not only simulate individual behaviors to form ABMs \citep{Park2023}, but also infer future disease dynamics based on population data and textual information of case reports \citep{Williams2023}. Usually, LLMs leverage large-scale unstructured and multimodal data to support epidemic modeling from a knowledge-driven perspective. Unlike traditional or deep learning approaches that rely on numerical mobility records, LLMs can extract and synthesize information from diverse textual and contextual sources related to human movement and disease dynamics. Furthermore, LLMs can process social media streams and news archives to infer real-time shifts in population mobility and public responses to intervention measures. By integrating these heterogeneous information layers, LLMs contribute to constructing a comprehensive data foundation that encompasses spatial characteristics, epidemiological trends, public health actions, and even genomic evidence, thereby enhancing the interpretability and situational awareness of epidemic analysis.

\begin{figure*}[htbp]
\centering
\includegraphics[width=\textwidth]{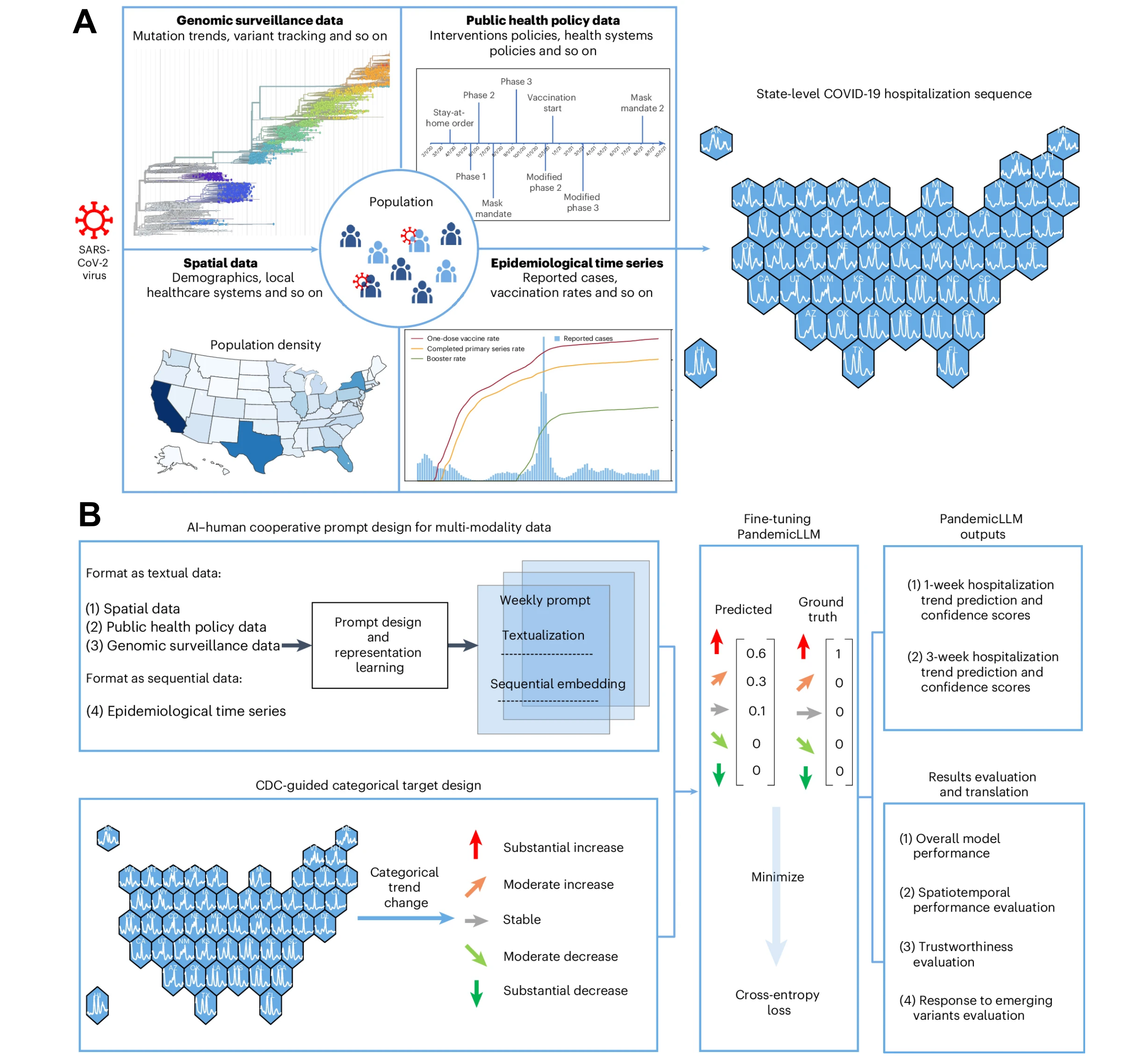}
\caption{Overview of pandemic data streams and pipeline in PandemicLLM.}

{\small\noindent
    \parbox{\textwidth}{
(A) The study utilizes a comprehensive dataset that brings together four key types of information related to the pandemic: spatial characteristics, epidemiological time series, public health interventions, and genomic surveillance. (B) In designing the PandemicLLM framework to forecast hospitalization trends during the pandemic, the problem is treated as an ordinal classification task. It adopts five outcome categories based on CDC guidelines: ranging from substantial decrease to substantial increase. To enable model training, multi-modal data are first reformatted into textual inputs through a collaborative prompt design involving both AI and human input. PandemicLLM is then fine-tuned using these text-based prompts and the associated prediction targets, focusing on both 1-week and 3-week timeframes. Careful evaluation procedures are applied to ensure the predictions are both accurate and reliable. This figure is a reproduction of Fig. 1 in Ref. \citep{Du2025}.
    }
}
\label{Fig.11}
\end{figure*}

An illustrative example is the PandemicLLM framework developed in \citep{Du2025}, which applies multimodal large language models to the field of infectious disease forecasting. In this approach, epidemic prediction problems are reframed as text-based reasoning tasks, allowing the model to integrate real-time, complex, and non-numerical information, such as public health policy texts, genomic surveillance data, and spatiotemporal epidemiological time-series data. The overview of this method is showed in \autoref{Fig.11}. It transforms traditionally heterogeneous data (e.g., epidemiological time-series, public health policy descriptions, genomic surveillance reports, and spatial-demographic indicators) into structured textual prompts. These prompts, typically around 300 words long, are carefully crafted through a cooperative design involving both human experts and AI systems. To handle the time-dependent nature of epidemiological data, the framework uses a recurrent neural network (GRU-based encoder) that converts sequential numerical data (like hospitalization rates) into dense embeddings. These embeddings are injected into the LLM’s input space via a special token mechanism, allowing the model to incorporate both temporal and textual information. Forecasting is formulated as a categorical classification problem, where the model predicts the future hospitalization trend of each U.S. state (for 1-week and 3-week horizons) into one of five categories: substantial decrease, moderate decrease, stable, moderate increase, or substantial increase. The LLM used (a fine-tuned version of Meta’s LLaMA2) is trained to generate the most likely category token based on the prompt. The model is evaluated across several time periods using not only conventional metrics like accuracy and mean squared error, but also probabilistic ones such as Brier score and ranked probability score, which better capture forecast uncertainty. Notably, three versions of PandemicLLM are trained on progressively larger datasets, and tested over extended periods without retraining, demonstrating the model's robustness and adaptability.

\section{Human Mobility and Epidemic Risk Management}\label{sec:6}

This section aims to illustrate how human mobility data can be used to identify risk sources, quantify infection pressure, and reveal transmission mechanisms, particularly during outbreaks driven by emerging pathogens. It also highlights studies focusing on early warning and outbreak detection.

\subsection{Epidemic Source Identification}\label{sec:6.1}

The rapid, large-scale, and diffuse movement of people can significantly amplify localized outbreaks into widespread epidemics through long-range travel, daily commuting, and socio-spatial congregation \citep{Perrotta2022}. Empirically, epidemics often originate from a few key locations or events, such as high-density transportation hubs or mass gatherings where large numbers of individuals converge \citep{Brockmann2013,Du2021a,Tatem2006}. Epidemic source identification is often associated with detecting highly mobile individuals who act as vectors, transmitting pathogens from high-risk regions to susceptible areas. This often involves analyzing travel records, mobile phone data, and other real-time mobility datasets (see \autoref{sec:2}) with epidemic models to locate the origins of outbreaks and trace how pathogens propagate across space and populations over time. Particularly group-level models (e.g., compartmental models, metapopulation models, and mobility network models, see \autoref{sec:5.1} to \autoref{sec:5.3.1}) are more effective (also computational efficient) for analyzing aggregated population-level transmission patterns and the large-scale spatial spread, making them play critical roles in risk assessment and intervention design (e.g., infection pressure estimation), while individual-level epidemic models (e.g., contact network models and agent-based models) are suitable for identifying infection origins and identifying super-spreaders due to their ability to depict individual heterogeneity (see \autoref{sec:5.3.2} and \autoref{sec:5.4}).

Empirical evidence from recent pandemics such as COVID-19 and influenza has been shown to correlate strongly with international and domestic human mobility patterns, as these movements contribute introducing infections to new, previously unaffected regions \citep{Poletto2013}. \autoref{Fig.12} (A) illustrates that nationwide outbreaks of infectious diseases often begin in a small number of early-affected regions that serve as risk sources \citep{Kim2019}. These initial outbreaks gradually trigger a chain of transmissions across the country, driven by human mobility and social interactions. While the overall epidemic may appear as a single nationwide curve, it can be decomposed into distinct regional timelines, each reflecting localized transmission dynamics. In this process, infections spread from the source regions to others via observable population movements and hidden social contact routes, highlighting the importance of early detection and intervention at the source to contain further spread. By tracking epidemic sources, public health officials can predict and manage the spatial spread of an epidemic, especially when mobility data is combined with real-time health surveillance \citep{Petersen2020}. This combination of mobility and individual health status from epidemic sources allows us to forecast where future outbreaks are most likely to occur and direct resources to these high-risk areas \citep{Germann2006,Kraemer2017,Poletto2013}.

Taking mobility trajectory (see \autoref{sec:3.1}) as an example, the use of smartphone and social media apps has demonstrated the potential of human mobility to assist in contact tracing efforts and epidemic source identification during pandemics \citep{Hernandez-Orallo2020}. Normally, these apps use Bluetooth technology to detect proximity between individuals and inform them if they have been in contact with someone who tested positive for the virus. Studies have shown that digital contact tracing systems can significantly reduce the time lag between an individual becoming infected and being isolated, thus curbing the spread of infections by identifying high-risk individuals faster \citep{Daniore2021,Nabeel2022,Tsvyatkova2022}. \autoref{Fig.12} (B) demonstrates the process of identifying risk sources based on trajectory data and inference methods, typically, machine learning models. The inputs are a series of GPS trajectories of individuals that have visited the outbreak origin (blue circle) and trajectories of other, unaffected individuals (grey, dashed lines). Using these data, people can infer the location and timing of an outbreak by identifying instances where infected individuals were in close proximity. The most prominent cluster is then identified as the estimated outbreak location through inference methods \citep{Lu2019,Schlosser2021}.

Study \citep{Jia2020} develops a spatiotemporal risk source model that utilizes population flow data (which operationalizes the risk emanating from epidemic epicenters) not only to forecast the distribution of confirmed cases but also to identify regions with a high risk of transmission at an early stage. The model yields a benchmark trend and an index for assessing the risk of community transmission of epidemics over time for different locations, and the effect of outflow on infection by using the following multiplicative exponential model:
\begin{equation}
{y_i} = c\prod\limits_{j = 1}^m {{e^{{\beta _j}{x_{ji}}}}{e^{\mathop \sum \limits_{k = 1}^n {\lambda _k}{I_{ik}}}}}
\text{,}
\label{eq:30}
\end{equation}
where $y_i$ is the number of cumulative (or daily) confirmed cases in location $i$; ${x_{1i}}$ is the cumulative population outflow from the epidemic source to location $i$; ${x_{2i}}$ is the economic indicator; ${x_{3i}}$ is the population size of location $i$; $m$ is the number of variables included, here $m = 3$; both $c$ and ${\beta _j}$ are parameters to be estimated. Besides, ${\lambda _k}$ is the fixed effect for location $k$; $n$ denotes the number of locations considered in the analysis; ${I_{ik}}$ is a dummy for location $i$ and ${I_{ik}} = 1$, if $i \in k$ (location $i$ belongs to $k$), otherwise ${I_{ik}} = 0$. This model leverages observed population flow data to operationalize the risk emanating from the epidemic source. It makes no assumption regarding travel patterns or practical distance effects, allows for nonlinear estimations, generates a non-arbitrary, source-linked risk score, and is easily adapted to other empirical contexts. \autoref{Fig.12} (C) demonstrates the predictive model based on population outflow. The left subgraph represents the outflow population from the epidemic center of COVID-19 in China (Wuhan) along with time, while the right subgraph represents the risk scores over time, providing a dynamic picture of shifting transmission risks in different prefectures in China.

Identifying epidemic sources as early as possible is essential for effective disease control and mitigation, as it allows public health authorities to implement timely and targeted interventions \citep{Kogan2021}. However, privacy concerns and public reluctance to adopt source identifying technologies remain major barriers to effective implementation. For instance, studies have shown that voluntary participation in digital contact tracing applications is often significantly lower than expected, mainly due to fears of surveillance and data misuse \citep{Fox2024}. These challenges undermine the timely identification of infection sources, which is essential for early intervention and targeted response. 

To overcome these limitations, future efforts should focus on improving the accuracy, reliability, and public acceptance of source identifying systems. By integrating multiple types of data, including multi-source human mobility traces, social media activity, environmental factors, and demographic information, researchers can build more comprehensive models for outbreak detection. In addition, advances in machine learning enhance the predictive performance of these epidemic source identification models, allowing for more adaptive and real-time epidemic management strategies \citep{Schwabe2021,Zeng2021}.

\subsection{Epidemic Risk Models of Infection Pressure}\label{sec:6.2}

The effectiveness of infectious disease response and containment can be significantly improved by concentrating healthcare efforts and control measures in areas at highest risk of new outbreaks. Calculating infection pressure is a quantitative method for evaluating the level of risk in specific areas. Infection pressure refers to the force exerted by pathogens' presence and transmission potential within a population, which influences the likelihood and degree of epidemic spread \citep{Yang2024}. It is determined by factors such as population density, mobility patterns, contact rates, and the characteristics of infectious disease. High infection pressure often arises in areas with large, mobile populations or frequent interactions, such as urban centers, transportation hubs, or regions with high rates of human mobility. Most studies calculate infection pressure based on mobility networks constructed by population flows between locations, and the data sources of population mobility often include public transportation records, cellular signaling data, and satellite positioning information. Some machine learning models have also applied to estimate infection pressure by integrating diverse human mobility data \citep{Yang2024}. However, such approaches inevitably involve a trade-off among data quality, interpretability, and accuracy.

Previous studies have demonstrated that infection pressure, calculated by mobility flows or contact rates, can serve as a valuable proxy for identifying early high-risk areas \citep{Du2020a,Jia2020}. For instance,  study \citep{Bai2022} analyzed \citep{Bogoch2020,Wu2020c}the travel-related risks associated with the SARS-CoV-2 Omicron variant in China, demonstrating that mobility data from air and rail travel were essential for estimating the importation and exportation risks of emerging variants, particularly in highly connected urban regions. Other studies leveraged air passenger itinerary data to estimate global infection pressure from high-risk Chinese cities, showing a strong correlation between air travel volumes and case importation risk, particularly before international travel restrictions were enacted . These models above calculate risk by population inflow rather than local population size, highlighting the directional and dynamic nature of infection risk propagation \citep{Wells2020}. They offer operational value for real-time risk assessment, early warning, and resource allocation, especially when combined with global travel data and local health system vulnerability indices \citep{Gilbert2020}.

\begin{figure*}[htbp]
\centering
\includegraphics[width=\textwidth]{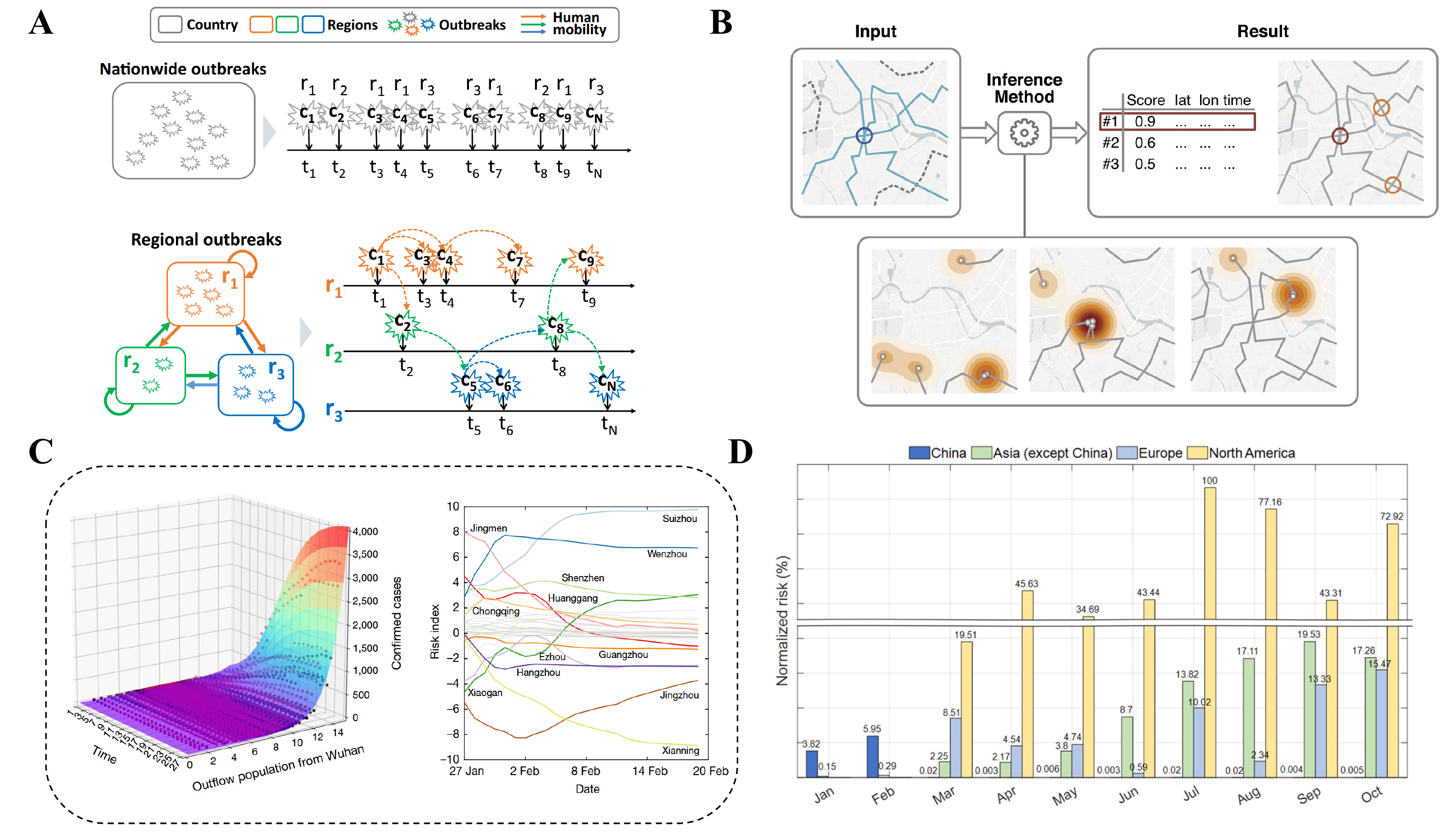}
\caption{Human mobility and epidemic risk assessment. }

{\small\noindent
    \parbox{\textwidth}{
(A) National outbreaks of epidemics often originate from a limited number of initial risk sources, with each outbreak event characterized by a contagion $c_i$ occurring at time $t$ in region $r_k$. (B) The method to infer the outbreak occurrence via human mobility data. The input consists of GPS trajectories from individuals who have visited the outbreak origin (blue circle), along with other unaffected individuals (grey dashed lines). The method seeks to infer both the location and timing of the outbreak using these trajectory data. To achieve this, the algorithm detects instances where individuals were in close proximity by identifying local maxima in their spatial distribution. The most prominent maximum is then determined as the estimated location of the outbreak. (C) Risk index calculation based on the outflow population from epidemic sources. In the left graph, the surface illustrates the fitted performance of the epidemic model against the cumulative population outflow from Wuhan, while the right graph shows the temporal evolution of \citep{Jia2020}the epidemic risk index across major cities in China. (D) The country-specific normalized risk (\%) of case importation from the top 13 countries to South Korea from January to October 2020. Fig. 12 (A), Fig. 12 (B), Fig. 12 (C) and Fig. 12 (D) are from Fig. 1 (A/B) in Ref. \citep{Kim2019}, Fig. 1 in Ref. \citep{Schlosser2021}, Fig. 3 in Ref. \citep{Jia2020} and Fig. 3 in Ref. \citep{Lee2021}, respectively.
    }
}
\label{Fig.12}
\end{figure*}

Generally, infection pressure integrates both the intensity of human mobility and the prevalence of infection at the origin, formalized as the product of population flow and incidence rate \citep{Brockmann2013,Wu2020c}. This indicator reflects not only the probability of seeding infections in target areas but also the temporal synchronization between mobility peaks and epidemic surges \citep{Badr2020}. As previously mentioned, it is significant to assess the infection pressure of importation of the virus from a risk source and its potential impact on the local transmission of the disease in regions impacted by the ongoing pandemic \citep{Brockmann2013,Nouvellet2021,Pappalardo2015}. Study \citep{Lee2021} focuses on the country-specific importation risk based on the number of international travelers, confirmed cases in the originating countries, and the population of the originating countries, proposing the risk assessment model based on Infection Pressure in month $t$ and originating country $c$ ($I{P_{c,t}}$):
\begin{equation}
I{P_{c,t}} = \frac{{{I_{c,t}}}}{{po{p_c}}}{T_{c,t}}
\text{,}
\label{eq:31}
\end{equation}
where ${I_{c,t}}$ stands for the monthly confirmed cases in month $t$ and originating country $c$. The population-adjusted density of infectious travelers was obtained by ${I_{c,t}}$ dividing its population $po{p_c}$ of country $c$. ${T_{c,t}}$ represents the number of passengers traveling from country $c$ in a month $t$. By applying the infection pressure’s formula, \autoref{Fig.12} (D) shows the normalized country-specific risk of case importation from the top 13 countries to South Korea from January to October 2020.

At the early stage of the outbreak, by constructing infection pressure indicators through real-time analysis of big data on population mobility, it is possible to predict the potential infection risks in various regions, assist the government in implementing regional graded control, and formulate refined prevention and control strategies to reduce the impact on the economy and society . In this process, compartmental models integrated human mobility data are widely used \citep{Chinazzi2020,Kraemer2020a} to estimate infection pressure. For instance, study \citep{Wu2020c} use data on flight bookings and human mobility data in China to predict the infection pressure of the COVID-19 virus and accounted for the effect of the Wuhan quarantine. Considering the infection pressure $z(t)$, the SEIR model can be extended as:
\begin{equation}
\left\{ {\begin{array}{*{20}{l}}
{\frac{{{\rm{d}}S(t)}}{{{\rm{d}}t}} =  - \frac{{S(t)}}{N}\left( {\frac{{{R_0}}}{{{D_I}}}I(t) + z(t)} \right) + {L_{I,W}} + {L_{C,W}}(t) - \left( {\frac{{{L_{W,I}}}}{N} + \frac{{{L_{W,C}}(t)}}{N}} \right)S(t)}\\
{\frac{{{\rm{d}}E(t)}}{{{\rm{d}}t}} = \frac{{S(t)}}{N}\left( {\frac{{{R_0}}}{{{D_I}}}I(t) + z(t)} \right) - \frac{{E(t)}}{{{D_E}}} - \left( {\frac{{{L_{W,I}}}}{N} + \frac{{{L_{W,C}}(t)}}{N}} \right)E(t)}\\
{\frac{{{\rm{d}}I(t)}}{{{\rm{d}}t}} = \frac{{E(t)}}{{{D_F}}} - \frac{{I(t)}}{{{D_I}}} - \left( {\frac{{{L_{W,I}}}}{N} + \frac{{{L_{W,C}}(t)}}{N}} \right)I(t)}
\end{array}} \right.
\text{,}
\label{eq:32}
\end{equation}
where $S(t)$, $E(t)$, $I(t)$, and $R(t)$ are the number of susceptible, exposure, infectious, and recovered individuals at the time $t$; ${D_E}$ and ${D_I}$ are the mean exposure and contagious period; ${R_0}$ is the basic reproductive number; ${L_{W,I}}$ and ${L_{W,I}}$ are the international outbound and inbound air passengers, while ${L_{W,C}}(t)$ and ${L_{C,W}}(t)$ are the daily numbers of all domestic outbound and inbound travelers, respectively. In \citep{Wu2020c}, researchers estimate the outbreak size of COVID-19 thus far in Wuhan and the probable extent of disease spread to other cities domestically. The results clearly indicate that if the pandemic had spread rapidly everywhere at the same rate as in Wuhan, outbreaks would have quickly emerged in many major Chinese cities.

Overall, tracking human mobility is essential for quantifying infection pressure, as it helps identify where a disease is likely to spread and assess the intensity of transmission dynamics across different regions. Moreover, understanding infection pressure is crucial for implementing targeted interventions, as it highlights areas where containment measures, such as social distancing, vaccination, or travel restrictions, should be prioritized to reduce the risk of outbreaks.

\subsection{Network-Based Epidemic Risk Assessment}\label{sec:6.3}
Over the past decades, extensive empirical evidence has demonstrated that the spread of infectious diseases relies heavily on contacts structured by the network topology (see \autoref{sec:5.3}) \citep{Brockmann2013,Pastor-Satorras2015,Zhang2025a}. These interactions occur not only across mobility networks between locations but also within contact networks among individuals. Therefore, analyzing epidemic dynamics through the lens of network theory has become a critical approach for assessing epidemic risk.

One representative study that demonstrates network-based epidemic risk assessment is presented in \citep{Bengtsson2015}, which integrated human mobility, derived from mobile phone records into a mobility network framework to estimate spatial variations in infectious risk. Mobility network ${M^{phone}}$ with elements ${m_{ij}}^{phone}$, indicates the average daily proportion of mobile phones relocating from area $i$ to $j$, comparing their last registered location on the day $t$ with their previous registered location on the day $t - 1$, and then they propose the infection pressure ${P_j}(t)$, in which ${c_i}(t)$ is the number of reported cases in study area $i$ on day $t$:
\begin{equation}
{P_j}(t) = \sum\limits_{i,i \ne j}^n {\left[ {m_{ij}^{phone}\sum\limits_{k = 1}^7 {{c_i}} (t - k)} \right]}
\text{.}
\label{eq:33}
\end{equation}

The findings calculated by ${P_j}(t)$ indicate that the probability of an epidemic occurring in a specific region and the initial severity of local outbreaks could have been predicted during the initial stages of the epidemic by utilizing case reports and the mobility network models.

Another representative study \citep{Zhang2020a} demonstrates how epidemic risk can be accessed via combining contact networks with compartmental models. Each epidemiological compartments is subdivided into discrete 5-year age groups (e.g., 0–4, 5–9, …, 60–64, 65+ years old), allowing for age-structured transmission dynamics. Susceptible individuals are exposed to an age-specific force of infection regulated by the average number of contacts per day that individuals in a given age group have with members of all other age groups. The epidemic transmission process on contact networks can be simulated by the following differential equations:
\begin{equation}
\begin{array}{l}
{{\dot S}_l} =  - \beta \sum\limits_{j = 1}^n {{M_{ij}}} \frac{{{I_j}}}{{{N_j}}}{\sigma _i}{S_i}\\
{{\dot I}_\iota } = \beta \sum\limits_{j = 1}^n {{M_{ij}}} \frac{{{I_j}}}{{{N_j}}}{\sigma _i}{S_i} - \gamma {I_j}\\
{{\dot R}_l} = \gamma {I_j}
\end{array}
\text{,}
\label{eq:34}
\end{equation}
where $n$ is the total number of age classes; ${S_i}$, ${I_i}$, and ${R_i}$ are the number of susceptible, infectious, and recovered individuals in age group $i$; ${N_i}$ denotes the total number of individuals in age class   (e.g., ${N_i} = {S_i} + {I_i} + {R_i}$); ${\sigma _i}$ is the susceptibility to infection of individuals in age group $i$; $beta$ and $\gamma$ represent the transmission and recovery rate, respectively. The contact network, ${M_{ij}}$, is the average number of contacts between individuals in age group $i$ with individuals in age group $j$. In \citep{Zhang2020a}, the matrix ${M_{ij}}$ represents the contact matrix for regular weekdays and the COVID-19 outbreak period, as estimated from social survey in specific cities.

As empirically demonstrated during COVID-19, the targeted removal of high-traffic airport nodes (modeled as node or edge removal) significantly reduced global transmission risk \citep{Li2020d,Perofsky2024,Zhong2021}. For instance, effective interventions targeting key nodes of transmission can prevent localized outbreaks from evolving into widespread epidemics, thereby minimizing the impact on public health and economies \citep{Perofsky2024}. During pandemics, the intrinsic network properties of mobility and contact networks can themselves serve as indicators for risk assessment, thereby supporting the design of more effective intervention strategies. Study \citep{Boguna2003,Bucur2020,DeArruda2014,Wen2017} quantified several node centrality metrics that govern outbreak scalability, as summarized in \autoref{tab.4}.

\begin{table}[htbp]
  \caption{Epidemiological roles of network centrality measurements.}

  \label{tab.4}

  \begin{tabular}{
    >{\raggedright\arraybackslash}p{3cm}
    >{\raggedright\arraybackslash}p{3cm}
    >{\raggedright\arraybackslash}p{4cm}
    >{\raggedright\arraybackslash}p{4cm}
  }
    \toprule
    \textbf{Network Centrality Measurement} &
    \textbf{Definition} &
    \textbf{Epidemiological Role in Mobility Network} &
    \textbf{Epidemiological Role in Contact Network} \\
    \midrule

    Degree \citep{Boguna2003} &
    Number of direct connections a node has. &
    High-degree nodes serve as hubs connecting many locations; controlling them (e.g., airport closures) can limit long-distance spread. &
    Individuals with many contacts are easier to be infected and are potential super-spreaders; interventions targeting them reduce transmission. \\
    \addlinespace

    Betweenness Centrality \citep{Wen2017} &
    Measures how often a node lies on shortest paths between other nodes. &
    Nodes bridging communities or cities; removing/controlling them can slow inter-community spread. &
    Individuals connecting otherwise separate groups; targeting them can contain outbreaks between communities. \\
    \addlinespace

    Closeness Centrality \citep{DeArruda2014} &
    Average shortest path length from a node to others. &
    Nodes with rapid access to many locations; controlling them can delay epidemic peaks. &
    Individuals who can quickly reach many others; interventions can slow outbreak propagation. \\
    \addlinespace

    Eigenvector Centrality \citep{Bucur2020} &
    Measures influence of a node considering neighbors’ importance. &
    Locations connected to influential hubs; controlling these amplifies impact on overall network. &
    Individuals connected to highly connected people; interventions here have cascading effects. \\
    \addlinespace

    Coreness \citep{Dorogovtsev2006,Kitsak2010,Kong2019} &
    Coreness of a node is defined as the largest value of $k$ for which the node belongs to the $k$-core of the network. &
    Core locations represent the most interconnected parts of mobility networks; outbreaks starting here can rapidly percolate through the entire network. &
    Individuals in the network core sustain persistent transmission chains; targeting these densely connected cores can effectively dismantle epidemic persistence and prevent large-scale spreading. \\
    \bottomrule
  \end{tabular}
\end{table}

\section{Human Mobility and Response Strategies Design}\label{sec:7}

This section covers intervention strategies that incorporate mobility restrictions. It also reviews evaluation methods and models used to assess the effectiveness of these interventions, including analysis of the relationship between policy stringency and health outcomes, as well as insights drawn from natural experiments.

\subsection{Disruption of Epidemic Transmission Pathways}\label{sec:7.1}
Disrupting epidemic transmission pathways involves strategically cutting off the key routes that facilitate pathogen spread across populations \citep{2023g,JiaweiFeng2025,Yu2020}. Leveraging multi-source human mobility data (see \autoref{sec:2}), such as public transportation records for structured travel flows, cellular signaling data for broad-scale movements, and satellite positioning information for real-time location patterns, it is possible to reconstruct and target these pathways at various resolutions, from global dispersal to neighborhood-level interactions.

Network-based models play a key role in this process \citep{Chinazzi2020}. Within mobility networks (see \autoref{sec:3.2} \& \autoref{sec:5.3.1}), locations are represented as nodes and population flow as weighted edges to quantify infection risks along with high-traffic routes \citep{Balcan2009}. In parallel, contact networks (see \autoref{sec:3.3} \& \autoref{sec:5.3.2}) reveal individual-level interpersonal links that govern transmission through close physical proximity and social interaction \citep{Salathe2010}. Integrating these network representations into compartmental models allows for simulating how interventions like border closures or route diversions can block interregional transmission chains \citep{Kostandova2024a}. Advancements in machine learning (see \autoref{sec:5.5}) enable the detection of hidden vulnerabilities by analyzing mobility fluctuations against outbreak data, pinpointing pathways where minor disruptions can significantly lower reproduction numbers \citep{Du2025}. Meanwhile, agent-based models (see \autoref{sec:5.4}) can further simulate individual-level responses, such as avoiding crowded paths, to predict how targeted measures can preemptively dismantle transmission networks \citep{Hinch2021}.

This integration of epidemic surveillance data, mobility analytics, and computational modeling helps public health authorities identify and interrupt super-spreader events and cascading transmission routes, thereby preventing further epidemic escalation through targeted pathway disruption \citep{Pastor-Satorras2015}. As shown in \autoref{Fig.13} (A), imposing targeted human mobility restrictions has been proved effective for disrupting epidemic transmission pathways during COVID-19 in Shanghai \citep{Zhang2023e}. Besides, protecting vulnerable populations (e.g., older adults, immunocompromised individuals, and residents of socioeconomically deprived areas) remains a cornerstone of effective disruption for epidemic transmission \citep{Jamison2013}. 

As evidenced by mobility interventions across diverse geopolitical contexts, the effectiveness of node- or edge-targeting strategies is profoundly shaped by local community structures and the phased deployment of policies \citep{Liu2022c}. In densely connected metropolitan areas, multi-stage interventions, beginning with the isolation of high-traffic hubs followed by the containment of peripheral sub-networks, tend to yield the most substantial reductions in transmission \citep{Maier2020}. Conversely, in sparsely populated or spatially fragmented regions, community-based containment and travel corridor management often prove more effective \citep{Chang2024}. Beyond structural factors, sociocultural norms, economic dependencies, and institutional capacities critically influence intervention outcomes. For instance, regions with strong public compliance and transparent risk communication exhibit faster network fragmentation and lower effective reproduction numbers \citep{Tan2021a}, while inconsistent policy timing or poor inter-jurisdictional coordination frequently results in premature reopening and epidemic resurgence \citep{Stone2007}. These observations underscore that the disruption of epidemic transmission pathways depends not only on the optimization of network structures but also on adaptive governance and behavioral responsiveness at the local level. Moreover, challenges remain in balancing the need for control with socio-economic considerations when disrupting transmission pathways of epidemics \citep{Das2024,Pardo-Araujo2023}.

\subsection{Mobility-Oriented Non-Pharmaceutical Interventions}\label{sec:7.2}

Mobility-oriented Non-Pharmaceutical Interventions (NPIs) are a set of strategies that aim to modify human movement patterns with the objective of reducing exposure and transmission risks, independent of medical treatments \citep{2021j,Du2022b,Lai2020,Nouvellet2021,Perra2021,Yang2021b}. These interventions can be tailored for effective epidemic control by drawing from datasets such as public transportation records, cellular signaling, GPS, and IoT-based location tracking (see \autoref{sec:2}). The timely implementation of interventions, such as quarantine and isolation, plays a key role in limiting transmission and minimizing socioeconomic disruptions \citep{Chinazzi2020,Riley2003}. For instance, mobility network models (see \autoref{sec:5.3.1}) are frequently used to evaluate NPI impacts by quantifying changes in network connectivity and flow metrics, demonstrating how policies like travel bans fragment transmission routes \citep{2023f,Xue2021}. By fusing these models with human mobility data, mobility-oriented NPIs evolve from broad mandates to adaptive strategies that optimize health outcomes while mitigating socioeconomic disruptions, ensuring sustainable control of epidemic dynamics. 

\begin{figure*}[htbp]
\centering
\includegraphics[width=\textwidth]{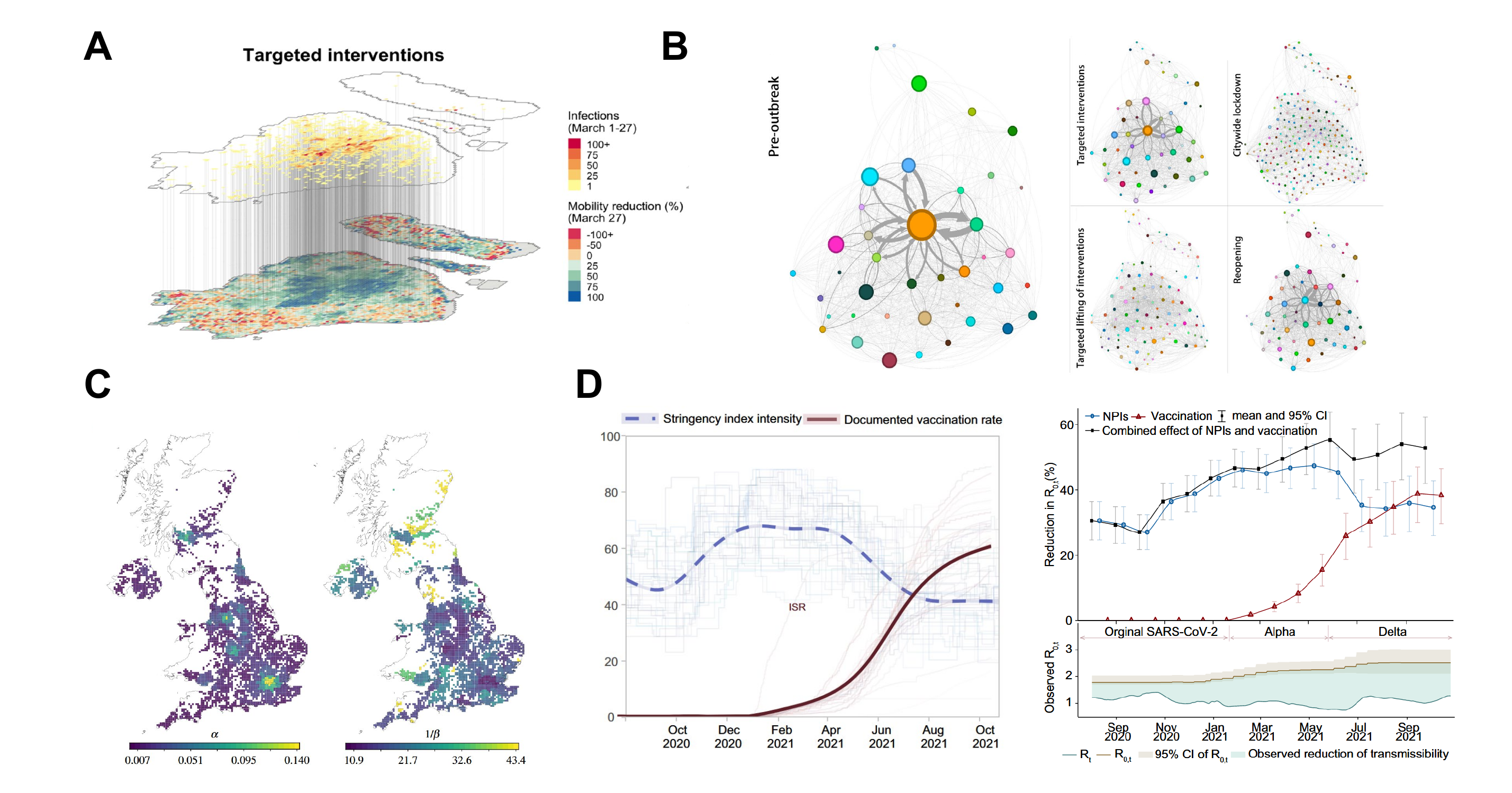}
\caption{Policy response under pandemics.}

{\small\noindent
    \parbox{\textwidth}{
(A) The geographic distribution of infections and mobility reduction during the targeted interventions phase in Shanghai. The upper map shows the number of infections at the grid level, while the lower map shows the reduction in mobility. (B) The change of mobility network structure in the context of pandemics. During the phases of pre-outbreak, targeted intervention, and reopening, the mobility network in Shanghai exhibits a clear central structure, where peripheral nodes are strongly connected to core nodes representing major urban centers. In contrast, during the phases of targeted lifting intervention and citywide lockdown, the network loses its central hubs, and population movements become largely confined within local communities. (C) Regional differences in the response to the lockdown in UK. The shock amplitude $\alpha $, representing the degree of change in human mobility, is higher in urban centers and lower in rural areas, whereas the recovery timescale $1/\beta $, indicating the return of mobility to pre-lockdown levels, exhibits the opposite pattern. (D) The left graph demonstrates the relationship between the stringency index of “lockdown” style NPIs and the documented vaccination rate across 31 countries. The documented vaccination rate refers to the proportion of the total population who were fully vaccinated in each country. The right graph shows that the key role of NPIs in epidemic intervention during the early outbreaks. The overall monthly effects of  interventions on reducing ${R_{0,t}}$ across 31 countries are presented with mean and 95\% CI. The total effect of NPIs presented here is the effect of NPIs alone plus their interaction effect with vaccination, and the total effect of vaccination shown is the impact of vaccination alone plus its interaction effect with NPIs. In the bottom panel, the light blue area between ${R_{0,t}}$ (instantaneous basic reproduction number) and ${R_t}$ (instantaneous reproduction number) illustrates the observed reduction of COVID-19 transmissibility. Fig. 13 (A) and Fig. 13 (B) are from Fig. 3 (A) and Fig. 2 in Ref. \citep{Zhang2023e}, while Fig. 13 (C) is from Fig. 5 in Ref. \citep{Schindler2023}. Fig. 13 (D) is from Fig. 1 (b) and Fig 2 in Ref. \citep{Ge2022}.
    }
}
\label{Fig.13}
\end{figure*}

During the COVID-19 pandemic, the government-imposed measures to control epidemic spread have altered human behavior and mobility patterns in an unprecedented way \citep{Yang2023}. At the beginning of the COVID-19 pandemic, Stringency Index (SI) was proposed the to quantify the intensity of NPIs \citep{Ma2021}. A study based on 113 countries showed a strong correlation between SI and the infection rate: for every 10\% increase in SI, the mortality rate decreased by approx. 6\% \citep{Hale2021a}. In addition, some study focused on the timeliness of cutting off the transmission pathways explored the quantitative relationship between reaching a high SI level earlier and reaching the peak of the epidemic earlier: from the initiation of response measures and the reporting of the first case, for every day earlier the high SI level is reached, the daily peak of new cases can be reached 0.44 days and 0.65 days earlier, respectively \citep{Ma2021}. For instance, a study based on barangay-level COVID-19 data in the Philippines found that the correlation between SI and the number of cases was affected by population density: when population density increased from 26,903 per km2 to 44,290 per km2, the correlation coefficient  decreased by 20\% (from 0.70 to 0.56) \citep{Necesito2022}. \autoref{Fig.13} (B) demonstrates the evolution of community structures in Shanghai across five distinct stages with corresponding SI: the pre-outbreak period, the targeted intervention phase, the citywide lockdown phase, the targeted lifting of the intervention phase, and the reopening phase. Each node was found to correspond to a community, and the node's centre was found to coincide with the centroid of the community. The size of each node was found to be proportional to the community's area (number of cells). The width of the directed arrow was found to be proportional to the flows between communities. 

A related study \citep{Xue2021} analyzes the multiscale structure of human mobility in the United Kingdom using Facebook Movement data collected before and during the first COVID-19 lockdown. By constructing a pre-lockdown mobility network and applying multiscale community detection, the researchers identified hierarchical flow communities that better reflected actual mobility patterns than administrative divisions. The results show that during lockdown, mobility contracted toward fine-scale local communities already present in pre-lockdown data and gradually expanded back to coarser communities as restrictions were lifted. The temporal dynamics of this process were effectively modeled by a linear decay shock model, revealing regional variations in both the magnitude and recovery speed of the mobility disruption, as shown in \autoref{Fig.13} (C). Similarly, the research examining the lagged impacts of mobility changes (using Google COVID-19 Mobility Research Dataset) and environmental factors on COVID-19 waves \citep{Du2022b} in rural and urban India found that recovery of mobility to 99\% of pre-pandemic levels was associated with an increased relative risk of transmission during the Delta wave. This resurgence in movement, combined with reduced policy stringency and the emergence of the Delta variant, was identified as a major driver of the sharp transmission peak observed in April 2021. Moreover, the dominant drivers of transmission varied across urban, rural, and suburban settings, underscoring the spatial heterogeneity of epidemic responses and outcomes.

In practical epidemic prevention and control, strategies are not formulated in isolation; both the epidemic intensity in surrounding areas and the local vaccination coverage play key roles in shaping the implementation and adjustment of NPIs in a given region. For instance, study \citep{Ge2022} used an integrated stringency index of government interventions generated by the Oxford Covid-19 Government Response Tracker (OxCGRT) \citep{Hale2021} as a proxy to estimate the general restraint of lockdown style NPIs. Researchers employ a Bayesian inference model to quantify the dynamic effects of NPIs and vaccination—individually and jointly—on COVID-19 transmission across Europe from August 2020 to October 2021. Using comprehensive datasets covering epidemiological factors, virus variants, vaccination progress, and climate conditions, \autoref{Fig.13} (D) shows that the combined impact of NPIs and vaccination reduced the reproduction number by 53\% by October 2021, while NPIs and vaccination alone reduced transmission by 35\% and 38\%, respectively. The study also finds that NPI effectiveness was less sensitive to emerging variants compared with vaccination, and its relative impact declined by 12\% after May 2021 due to easing restrictions and vaccine rollout.

\section{Conclusion and Outlook}\label{sec:8}
Human mobility has become a cornerstone for better understanding, predicting, and mitigating the spread of infectious diseases \citep{Belman2024,Jia2020,Jusup2022,Kraemer2020a,Lai2020}. With the advancement of technologies, a diverse array of mobility data sources is now available, including social surveys, public transportation, cellular signaling, satellite positioning, IP and WiFi addresses, and IoT information. These data can be represented in various forms, such as trajectories, mobility networks, contact networks, and aggregated indices, providing flexible and scalable frameworks to model complex human interactions and movement patterns \citep{Alessandretti2022,Barreras2024}. Through these representations, researchers have uncovered valuable and quantifiable insights of correlations between population flow and epidemic transmission, offering new opportunities to develop more precise risk assessments and more responsive public health measures \citep{Germann2006,Kraemer2017,Poletto2013,Salje2021}.

A pandemic with a high $R_0$ value that has gained a foothold somewhere in the world cannot be stopped—no matter how good the data is—but that good data can help mitigate its impact \citep{Chen2008}. However, the integration of mobility data into a range of epidemic modelling paradigms, such as compartmental models, complex network models, agent-based simulations, and machine learning approaches, has significantly enhanced the resolution and realism for epidemiological forecasts and interventions \citep{Castillo-Chavez2016,Li2021d,Liu2023d,Sallah2017,Tan2021,Urabe2016,Wang2017,Zhang2023e}. For example, informed by near real-time mobility data, SIR and metapopulation models can simulate spatially resolved disease dynamics with higher fidelity, while agent-based simulations capture individual-level heterogeneity across different geographical regions \citep{Kogan2021}. Moreover, machine learning techniques are increasingly used to process large-scale mobility datasets and extract latent features that drive epidemic propagation, thereby strengthening early warning systems and supporting adaptive policy responses \citep{Du2025}.

Despite these advancements, several methodological limitations and challenges need to be addressed to maximize the effectiveness of mobility-informed epidemic modeling. One critical area is the presence of demographic and socioeconomic biases in mobility datasets. Human mobility and contact patterns, derived from novel digital sources like mobile phones, GPS, and digital platforms, often tend to disproportionately represent urban, affluent, or tech-savvy populations, while under-representing others, particularly rural, low-income, or digitally excluded \citep{Barreras2024,Haddad2022}. These biases might distort model outputs and lead to inequitable or ineffective public health interventions. Although techniques such as statistical re-weighting, data imputation, and demographic re-calibration have been proposed to correct for these disparities, the effectiveness and generalizability of these methods across different contexts and populations remain an active area for further investigation.

Additionally, data gaps can emerge during crises, where the influx of real-time data exceeds processing capacities or when critical data sources become unavailable due to infrastructure failures, leading to incomplete or delayed datasets. In such situations, data fusion approaches, which combine data from multiple sources (e.g., mobile data, transportation systems, social media, and healthcare records), offer a promising strategy to fill gaps and triangulate insights \citep{Du2025,2023g,Liu2023c}. However, integrating datasets from heterogeneous sources introduces new challenges in terms of data consistency, quality, and interoperability, necessitating the development of robust methodological approaches to ensure reliability.

The ethical and privacy concerns are always paramount when working with granular human mobility data, especially from personal devices. Location tracking and contact tracing systems can potentially expose sensitive information, contributing to surveillance risks or societal misuse \citep{Kostandova2024}. Although privacy-preserving techniques like differential privacy and data anonymization have been proposed to mitigate these concerns, their implementation in real-world epidemic modeling remains limited. Moving forward, it is imperative that epidemic modeling frameworks adopt transparent, ethical standards for data use. 

Addressing these limitations requires a multi-pronged, interdisciplinary approach \citep{Lu2025b}. First, an integrated approach that combines emerging digital data streams with traditional methods (e.g., household surveys) can improve coverage and enable cross-validation and extrapolation of real-world human mobility patterns. Second, enhancing multi-institutional data-sharing protocols across sectors, fostering collaborative research frameworks, and establishing standardized data processing approaches can help correct data bias to improve the accuracy, privacy and inclusivity of mobility-informed epidemic models. Moreover, the development of advanced statistical methods and validation techniques for handling missing or uncertain data will also enhance the robustness and credibility of model predictions \citep{2024b,2023a}. As artificial intelligence and machine learning continue to evolve, these technologies can be leveraged to identify hidden patterns and correlations in complex, large-scale mobility datasets, thereby improving model precision \citep{Du2025}. However, it is equally important that researchers and policymakers collaborate to establish clear ethical guidelines and privacy safeguards to ensure responsible use of human mobility data and sophisticated analyzing technology. 

Integrating human behavioral data into epidemic models still represents one of the most promising frontiers in infectious disease research, with immense potential for enhancing our understanding and control of infectious diseases. As the global community prepares for future public health crises, it is critical to continue investment in data infrastructure, analytic capability, and privacy-preserving innovation. In particular, strengthening the foundations for real-time mobility monitoring, digital contact tracing, and precise epidemic simulations in more inclusive, ethical, and actionable ways, will enhance our collective ability to respond swiftly and equitably to emerging threats.

\section*{Acknowledgments}

This work was supported by the National Natural Science Foundation of China (72025405, 72421002, 92467302, 72301285, 72474223), the Hunan Science and Technology Plan Project (2023JJ40685, 2024JJ6069), and the Major Program of Xiangjiang Laboratory (24XJJCYJ01001). The authors declare that they have no conflict of interest.

\section*{CRediT author statement}
Conceptualization: XL, HJY and FL; Data curation: XL, JWF, XQY, XYZ and SL; Methodology: XL, JWF, SJL, PH, SL, SQW and ZWD; Writing: JWF, XL, FL, PH, SJL, ZWD, XQY, YXL, XYZ, YB, XJD, WJM, HJY and SYT; Reviewing and Editing: all authors.

\bibliographystyle{elsarticle-num}    
\bibliography{Ref_update}

\end{document}